%% file: main_R.tex
\documentclass[12pt,3p,review,doublespacing]{elsarticle/elsarticle}
\journal{...}
\usepackage{sectsty}
\sectionfont{\color{blue}\MakeUppercase}
\subsectionfont{\color{blue}}

\fontfamily{cmss}\selectfont
\usepackage[scaled=0.99]{helvet}
\emergencystretch=\maxdimen
\graphicspath{{Figures}{Figures/Figure_01}{Figures/Figure_02}{Figures/Figure_03}{Figures/Figure_04}{Figures/Figure_05}{Figures/Figure_06}{Figures/Figure_07}{Figures/Figure_08}{Figures/Figure_09}{Figures/Figure_10}{Figures/Figure_11}{Figures/Figure_12}{Figures/Figure_13}}
\usepackage{hyperref}
 \hypersetup{
 	plainpages = false, 
	bookmarksopen = true,
	bookmarksnumbered = true,
	breaklinks = true,
	linktocpage,
	colorlinks = true,
	linkcolor = purple,
	urlcolor  = black,
	citecolor = purple,
	}
\usepackage{natbib}
\setcitestyle{round, square, number}
\usepackage{multicol}
\usepackage{multirow}
\usepackage{imakeidx}
\usepackage{nomencl}
\makenomenclature
\usepackage[xindy]{glossaries}
\makeglossaries
\usepackage{graphics}
\usepackage{graphicx}
\usepackage{epsf,psfrag}
\usepackage{epstopdf}
\usepackage[para]{threeparttable}
\usepackage{subfig}
\usepackage{epsfig}
\usepackage{pdflscape}
\usepackage{rotating}
\usepackage{lscape}
\usepackage{float}
\usepackage{stfloats}
\usepackage{textgreek}
\usepackage{longtable}
\usepackage{lscape}
\usepackage{capt-of}
\usepackage{comment}
\usepackage{tablefootnote}
\usepackage{footnote}
\usepackage{caption}
\usepackage[autopunct=true]{csquotes}
\usepackage{rotating}
\usepackage{txfonts}
\usepackage[normalem]{ulem}
\usepackage{resizegather}
\usepackage{siunitx}
\usepackage{setspace}
\usepackage{color,xcolor,soul}
\usepackage{amsfonts,amsmath,amssymb}
\usepackage{latexsym,array}
\usepackage{textcomp}

\newcommand{\RomanNumeralCaps}[1]
\linenumbers
\makesavenoteenv{tabular}
\makesavenoteenv{table}
\usepackage[left,displaymath, mathlines]{lineno}
 
\DeclareMathAlphabet{\mathpzc}{OT1}{pzc}{m}{it}
\def\bf{{\mbox{\boldmath $\rm f$}}}

\def\fig{Figure~}
\def\figs{Figures~}
\def\eqn{Eq.~}
\def\eqns{Eqs.~}
\def\tab{Table~}

\newcommand{\myvec}[1]{\mathbf{#1}}     
\def\tsc#1{\csdef{#1}{\textsc{\lowercase{#1}}\xspace}}
\tsc{WGM}
\tsc{QE}
\tsc{EP}
\tsc{PMS}
\tsc{BEC}
\tsc{DE}
\newcommand{\rev}[1]{{#1}}     
\newcommand{\ttxt}[1]{\textbf{\color{blue}\MakeUppercase{#1}}}     

\begin{document}

%
%
\setcounter{page}{1}
\begin{frontmatter} 
%
%
%
%
\title{\ttxt{Charge-dependent slip flow of ionic liquids through the non-uniform microfluidic device: pressure drop and electroviscous effects}}
%
%
%
%
\author[labela]{Jitendra {Dhakar}}
\author[labela]{Ram Prakash {Bharti}\corref{coradd}}\ead{rpbharti@iitr.ac.in}
\address[labela]{Complex Fluid Dynamics and Microfluidics (CFDM) Lab, Department of Chemical Engineering, Indian Institute of Technology Roorkee, Roorkee - 247667, Uttarakhand, INDIA}
%
%
\cortext[coradd]{\textit{Corresponding author. }}
%
\begin{abstract}
\fontsize{12}{18pt}\selectfont
This work investigates electroviscous effects in the presence of charge-dependent slip in steady pressure-driven laminar flow of a symmetric (1:1) electrolyte liquid through a uniformly charged slit contraction - expansion (4:1:4) microfluidic device. The mathematical model comprising the Poisson's, the Nernst-Planck, the Navier-Stokes, and the current continuity equations are solved numerically using the finite element method (FEM).  The flow fields (electrical potential, charge, induced electric field strength, pressure drop, and electroviscous correction factor) have been obtained and presented for the wide range of the governing parameters like inverse Debye length ($2\le K\le 20$), surface charge density ($4\le S\le 16$) and the slip length ($0\le B_0\le 0.20$) at fixed Schmidt number ($\mathit{Sc}=1000$) and low Reynolds number ($Re=0.01$).  
The flow fields have shown complex dependence on the governing parameters. The charge-dependent slip has further enhanced the complexity of the dependency in comparison to the no-slip condition. In presence of charge-dependent slip, the total electrical potential \rev{($|\Delta U|$) maximally increases by 78.68\%} and pressure drop \rev{($|\Delta P|$)} maximally \rev{decreases} by 63.42\%, relative to no-slip flow, over the ranges of conditions. The electroviscous correction factor ($Y = $ ratio of apparent to physical viscosity) increases by 33.58\% under the no-slip ($B_{\text{0}}=0$) condition. In contrast, the electroviscous correction factor ($Y$) increases maximally by 72.10\% for charge-dependent slip than that in the no-slip flow for the considered ranges of the conditions.  
A simple analytical model to estimate the pressure drop in the electroviscous flow \rev{has been} developed based on the Poiseuille flow in the individual uniform sections and pressure loss due to thin orifice\rev{. The model} overpredicts the pressure drop by 2 - 4\% \rev{from the numerical values}.
Finally, the predictive relations, depicting the functional dependence of the numerical results on the governing parameters, are presented for their practical use in the design and engineering of microfluidic devices.
\end{abstract}
\end{frontmatter}
%
\section{Introduction}
\label{sec:intro}
%
\noindent 
The importance of micro-electro-mechanical systems (MEMS) is continuously increasing because of their wide applications in the industrial science and engineering fields \citep{bhushan2007springer,li2008encyclopedia,lin2011microfluidics}. The `microfluidic' flow is different from the conventional large-scale `macrofluidic' flow as it depicts various features which are remarkably affected by the surfaces and interfaces. Amongst other factors, the surface charge and slip boundary condition \citep{churaev1984slippage,jing2015coupling,navier1827memorie,vinogradova1995drainage,pan2014study} on the wall of microfluidic device play an essential role in the transport of liquids.

\noindent
Electrokinetic phenomena evolve when solid surfaces (or materials and interfaces such as PDMS, glass) interact with electrolyte liquid \citep{Hunter2018,
Hunter2001,li2001electro,Schoch2005,Srinivasan2006,Delgado2007,Nakamura2011,
Somasundaran2015}. The charged surfaces attract counter-ions and repeal co-ions of electrolyte liquid. In the close vicinity of the surface, the counter-ions get attached to the surface due to the strong electrostatic force of attraction and form a rigid layer called an `immobile compact layer'. 
The compact \rev{(or Stern)} layer includes the `charged free' region (i.e., inner Helmholtz plane, IHP) containing excess counter-ions (and deficit of co-ions) near the charged surface, followed by the outer Helmholtz plane (OHP). 
Subsequently, a `diffusive layer' of ions forms away from the surface, where the electrostatic attraction force is weak, and the ions within the layer are mobile. The two (compact and diffusive) layers neutralize the charged surface in the liquid and are known as an electrical double layer (EDL). \rev{It implies quite a common assumption that the shear plane and the OHP are co-located.}
In EDL, the electrical potential ($\psi$) linearly decreases from the surface (actual thermodynamic potential $\psi_0$) to \rev{IHP, and again from IHP to OHP (Stern potential, $\psi_d$)}. \rev{Further, the elctrical potential} exponentially decays to zero in the diffuse layer.
Zeta (or electrokinetic) potential ($\zeta$) is defined as the potential at the shear plane (slip plane) forming the interface between the compact and diffuse layers of EDL and remains attached to the surface. 

\noindent 
When the pressure-driven flow approaches over (or through) such surfaces, the transport of ions in a diffusive layer generates a current known as `streaming current'. The accumulation of ions at the downstream end creates a potential difference between the upstream and downstream ends of the device called `streaming potential'. It drives counter-ions in EDL in the direction opposite to the pressure-driven flow and generates a current known as `conduction current'. In turn, additional hydrodynamic resistance is developed as the streaming potential exerts an extra body force on the charged liquid in EDL and induces an electro-osmotic back-flow that retards the primary pressure-driven flow. Consequently, the effect on pressure drop is the same as if liquid viscosity has increased, without electrokinetic effect, at the fixed volumetric flow rate. Therefore, the resulting increase in hydrodynamic resistance is known as the `electroviscous effect' \citep{Hunter2018,Atten1982}. 

\noindent
\rev{To the best of our knowledge,} the combined influences of inherent surface charge, charge-dependent slip, and geometrical features on microfluidic hydrodynamics are unexplored in the literature. The present work investigates electroviscous effects in the presence of the surface charge-dependent slip in the pressure-driven flow of electrolyte liquids through a uniformly charged \rev{slit} contraction-expansion microfluidic device. 
It constitutes a novel problem of intensifying microfluidic hydrodynamics by exploiting the intrinsic surface (surface charge and charge-dependent slip) and geometrical features (non-uniform geometry). 
\rev{The drag remarkably increases (and decreases) with increasing electroviscous (and slip) effects. The non-uniform geometries (like sudden contraction or expansion) further influence the drag.} 
\rev{Both charge-dependent slip and non-uniform geometry effects increase the electroviscous impact, i.e., retards the primary pressure-driven flow of liquid and increases the residence time for a fixed length of microchannel. The present results, thus, can be utilized to intensify the microfluidic transport processes, including mixing, diffusion, heat and mass transfer, reaction, etc.}
Further, the simple semi-analytical model presented in this work for easy determination of pressure drop is another novelty. The outcome of this work finds its significance in efficiently designing biomedical and related applications such as drug delivery, DNA sequencing, and biochemical analysis.
At this stage, it is informative to present a systematic review of the relevant literature to define the objectives and formulate the physical problem.
%
\section{Background literature}
\label{sec:2}
%
\noindent 
Over the decades, considerable research attention has been given to exploring the electroviscous influences in pressure-driven flow through microfluidic devices of various cross-sections and geometrical configurations for broader flow conditions. However, most experimental and numerical studies have accounted for the no-slip channel walls.
%
For instance, the first pioneering studies have explored the electroviscous effects in the no-slip Newtonian fluid flow through uniform slit \citep{burgreen1964electrokinetic} and cylindrical \citep{rice1965electrokinetic} microchannels. 
\rev{\citet{burgreen1964electrokinetic} obtained the analytical solution using the general theory of electrokinetics for small electrokinetic radius. \citet{rice1965electrokinetic} theoretically analyzed the electroviscous effects by invoking the Debye-Huckel (D-H) approximation for low zeta potential ($\zeta\le 25$ mV).}
These studies \citep{burgreen1964electrokinetic,rice1965electrokinetic} highlighted that, for a fixed $\zeta$ potential, the electroviscous effects (ratio of apparent to bulk viscosity) decrease with increasing electrokinetic width (i.e., product of characteristics length and inverse Debye length). 
%
%
\citet{levine1975theory} extended the work of \citet{rice1965electrokinetic} and solved the exact Poisson-Boltzmann equation (PBE) for high $\zeta$ potential. \citet{bowen1995electroviscous}, without invoking D-H approximation,  have shown that cation mobility considerably affected the electroviscous effects in a cylindrical microchannel. 

\noindent 
\citet{li2001electro} presented a broader discussion about the electroviscous impacts in pressure-driven liquid flow through microfluidic devices. The experimental and numerical study \citep{ren2001electro} has shown increased pressure drop ($\Delta P$), mainly due to electroviscous effects, in a rectangular microchannel for pure water and dilute aqueous ionic solutions. 
\citet{hsu2002electrokinetic} studied the electroviscous effects in an elliptical microchannel with the variation of aspect ratio and electrical boundary conditions (constant surface charge, constant $\zeta$ potential, and charge-regulated surface). 
\citet{chun2003electrokinetic} obtained an analytical solution of non-linear PBE to quantify electroviscous effects in slit microchannel flow. Their results depicted stronger influences of ionic concentration, $\zeta$ potential, and wall charge on the streaming potential and velocity profiles.
\citet{ren2004electroviscous} have developed a new theoretical model using the Nernst–Planck equation (NPE) to study the electroviscous effects on electrolytes flow in a slit microchannel. Another study \citep{chen2004developing} on the electroviscous effects in developing pressure-driven flow through parallel slit microchannel has shown that the streaming potential varies rapidly and becomes constant as it approaches fully-developed nature. 
\citet{stone2004engineering} briefly reviewed the electrokinetic flow of Newtonian fluids through microfluidic devices. \citet{brutin2005modeling} have modeled the surface-fluid electrokinetic coupling on the laminar flow in microtubes. They observed that the Poiseuille number ($Po=f\times Re$) is independent of the average velocity, even in the presence of EDL. Their model agrees well with experiments conducted at high surface potentials ($> 25$ mV) with microtubes (530 to 50 $\mu$m).
A featured article \citep{Delgado2007} has presented the progress of electrokinetics and recommended the applicable rules for measurements and interpretations of the electrokinetic (or $\zeta$) potential. 

\noindent 
Further, \citet{gong2013electrokinetic} have studied the electrokinetic flow in the capillary microchannel and proposed an approach to measure the streaming potential. \citet{hsu2016electrokinetics} have numerical\rev{ly} explored the electrokinetics in the silica channels using three EDL models like Gouy-Chapman (GC), Basic Stern (BS), and Viscoelectric (VE) models. 
\citet{kim2018analysis} have studied the flow through nanochannels using effective ion concentration and quantified the electroviscous effects in terms of a newly introduced parameter (ratio of $\zeta$ potential to D-H parameter). 
\citet{jing2018optimization} have performed the optimization analysis for electroviscous influences on the fluid flow through a fractal tree-like microfluidic device to obtain the minimum hydraulic resistance. They found that the surface charge strongly affected the optimal tree-like structure of the device and modified the well-accepted Murray’s law by increasing its complexity.  Recently, \citet{riad2020analysis} studied the multilayer electroviscous flow in a shear-driven charged slit microfluidic device. They found a strong influence of the surface charge on the moving interface and EDL thickness on the streaming potential and fluid flow. Above a threshold, streaming potential flow reverses the main shear-driven flow near the charged wall.

\noindent 
Subsequent rigorous studies have quantified the electroviscous effects in symmetric electrolyte flow through microchannels of non-uniform cross-section such as contraction-expansion rectangular/slit  \citep{davidson2007electroviscous,davidson2008electroviscous,Berry2011} and cylindrical  \citep{bharti2008steady,bharti2009electroviscous,davidson2010electroviscous} for the fixed volumetric flow rate. They have shown a stronger influence of governing parameters (Debye length and surface charge density) on flow characteristics (EDL potential, excess charge, pressure drop, and electroviscous correction factor) for fixed Reynolds and Schmidt numbers.  They also proposed a simple analytical model \citep{davidson2007electroviscous,bharti2008steady} based on the pressure drop in Poiseuille flow to predict the pressure drop in contraction-expansion microchannel by summing up the pressure drop in individual uniform upstream, contraction and downstream sections with an addition of extra pressure drop due to sudden contraction-expansion ($\Delta P = \Delta P_{\text{u}}+\Delta P_{\text{c}}+\Delta P_{\text{d}}+\Delta P_{\text{e}}$). The simpler models predicted the pressure drop within $\pm 5\%$ of their numerical results. 

\noindent
As discussed above, electroviscous effects in the no-slip flow have been explored thoroughly for wide-ranging conditions. 
In contrast, the surface features (like boundary slip, surface charge) play an essential role in the dynamics of microfluidic flows; limited efforts are devoted to understanding the corresponding influences in slip flow. 
For instance, \citet{navier1827memorie} has first introduced the concept of the boundary slip at the wall, which considered the relative movement of the solid and liquid surface boundary.  A relation has been proposed between the velocity in the tangential direction in the flow field proportional to the perpendicular velocity gradient to the boundary; for thin EDL cases, Navier slip boundary at the wall is realistic \citep{navier1827memorie}. 
The existing literature has accounted degree of slip length in the range of several nanometers to the tens of micrometers \citep{navier1827memorie,churaev1984slippage,vinogradova1995drainage,joly2006liquid,pan2014study,jing2015coupling}. 
Both theoretical and experimental studies have concluded that the surface charge affected the boundary slip. 
\citet{joly2006liquid} have developed a mathematical model to explore the effect of surface charge on boundary slip using molecular dynamics (MD) simulation and found that the higher surface charge density ($\sigma$) results in a lower slip length. 
\citet{wang2010flow} have investigated the electroviscous effects on the liquid slip flow in slit microchannels made of different materials. They concluded that the wall slip increased the flow-induced electric field and enhanced the electroviscous effects.
 \citet{jamaati2010pressure} have analyzed the electroviscous slip-flow in a planar microchannel by solving the non-linear PBE without invoking D-H approximation. They observed an increase in the induced voltage significantly with velocity slip at the wall.

\noindent \citet{jing2015coupling} have presented a comprehensive review of the surface charge and boundary slip at the solid-liquid interface and their combined effects on fluid drag. They further explored the electroviscous flow in a parallel–plate microchannel with high $\zeta$ potential and charge-dependent slip at the wall \citep{jing2015electroviscous}. They reported the reduction in fluid velocity, hence drag enhancement, with increasing surface charge and increased flow rate, thus reduced drag, with the boundary slip. 
\citet{jing2017non} studied the overlapping EDL induced electroviscous effects and surface charge-dependent slip effects in a fluid flow through a parallel-plate microchannel. They have shown decreasing trends of the electroviscous and the fluid drag with slip with increasing $\zeta$ potential for the large enough $\zeta$ potential.  
\citet{buren2018electroviscous,buren2019effects} explored the effects of surface charge and boundary slip on time-periodic pressure-driven flow and electrokinetic energy conversion in parallel-plate and cylindrical nanochannels. They found that the slip is dependent on the surface charge; a higher surface charge reduces the slip length. Surface charge-dependent slip increases the fluid velocity and energy conversion in the nanochannel than the no-slip condition.
%
Recently,  \citet{sen2020slip} have analyzed the electroviscous and charge-dependent slip effects in nanofluid flows. They have shown enhanced ionic conduction due to the slip condition. Even for a constant slip length, the velocity slip at the wall shows variation with the salt concentration, channel length, and electroviscous effects. 
More recently, \citet{dhakar2022b} presented preliminary results on the slip effects in ionic liquids flow through a contraction--expansion microfluidic device for limiting conditions.

\noindent
Furthermore, various molecular dynamics (MD) studies have attempted to understand the interplay between the surface charge and electrolytes flow. For instance, few studies  \citep{thomas2008reassessing,kumar2012slip,kannam2013fast} have shown the scale-dependent relationship between flow enhancement and slip length in nanochannels for pressure-driven flows. In contrast, \citet{celebi2018molecular}  suggests that the electroosmotic slip flow is independent of the channel height. 
However, other studies \citep{celebi2017electric,celebi2018surface} show a correlation between the slip length and surface charge density, i.e., slip length decreases with increasing surface charge that affects the liquid transport. 
Other studies \citep{rezaei2015surface,celebi2017electric,celebi2018surface} have also shown increasing viscosity with increasing surface charge, as the surface charge affects the alignment of water molecules and ions, thereby creating different orientations and forming new hydrogen bonds. 
On the other hand, continuum mechanics study \citep{davidson2007electroviscous} have shown the electroviscous flow through slit microchannel is the scale-dependent based on the calculations performed at various microchannel characteristic lengths ($W=100, 200, 500$ and $1000$ nm), bulk ion concentration ($n_0$) and surface charge density ($\sigma$). 
\rev{These continuum studies \citep{davidson2007electroviscous,davidson2008electroviscous,
bharti2008steady,bharti2009electroviscous,davidson2010electroviscous, Berry2011} have shown that the apparent or effective ($\mu_\text{eff}$)} viscosity increases with increasing surface charge due to additional resistance in the liquid flow \rev{imposed by the induced streaming potential in the} microfluidic device increases with the enhancement in the surface charge density.

\noindent
\rev{Even though both `viscoelectric' and `electroviscous' effects originate from the charged surfaces, they fundamentally differ significantly. For instance, the viscoelectric effect leads to a change in the physical viscosity of the fluid \citep{Hunter2018}. However, the latter (electroviscous effect) does not alter the physical viscosity of the liquid \citep{davidson2007electroviscous,davidson2008electroviscous,
bharti2008steady,bharti2009electroviscous,davidson2010electroviscous, Berry2011}, but the apparent viscosity is a purely theoretical quantification of the flow resistance induced by the streaming potential.}

\noindent
In summary, the above efforts have mainly explored the electroviscous effects for uniform geometries with or without boundary slip on the microfluidic device walls. Fewer attempts have accounted for the non-uniform geometries but with the no-slip boundary condition. To the best of our knowledge, none of the efforts are evident to explore the electroviscous effects in the presence of charge-dependent boundary slip in non-uniform geometries, which is the aim of the present study. 

\noindent
This article investigates the electroviscous effects in the presence of charge-dependent slip in the pressure-driven symmetric electrolytes flow through the slit contraction-expansion microfluidic device.  The mathematical model which governs the flow physics is solved numerically using the finite element method (FEM). 
The detailed results (like electrical potential, excess charge, induced field strength, pressure drop, and the electroviscous correction factor) have been obtained and presented in this work for the wide range of non-dimensional parameters (surface charge density, $4\le S\le 16$; Debye parameter, $2\le K\le 20$; and slip length, $0\le B_0\le 0.20$). 
%
\section{Physical and mathematical modelling}\label{sec:3}
\begin{figure}[h]
	\centering\includegraphics[width=1\linewidth]{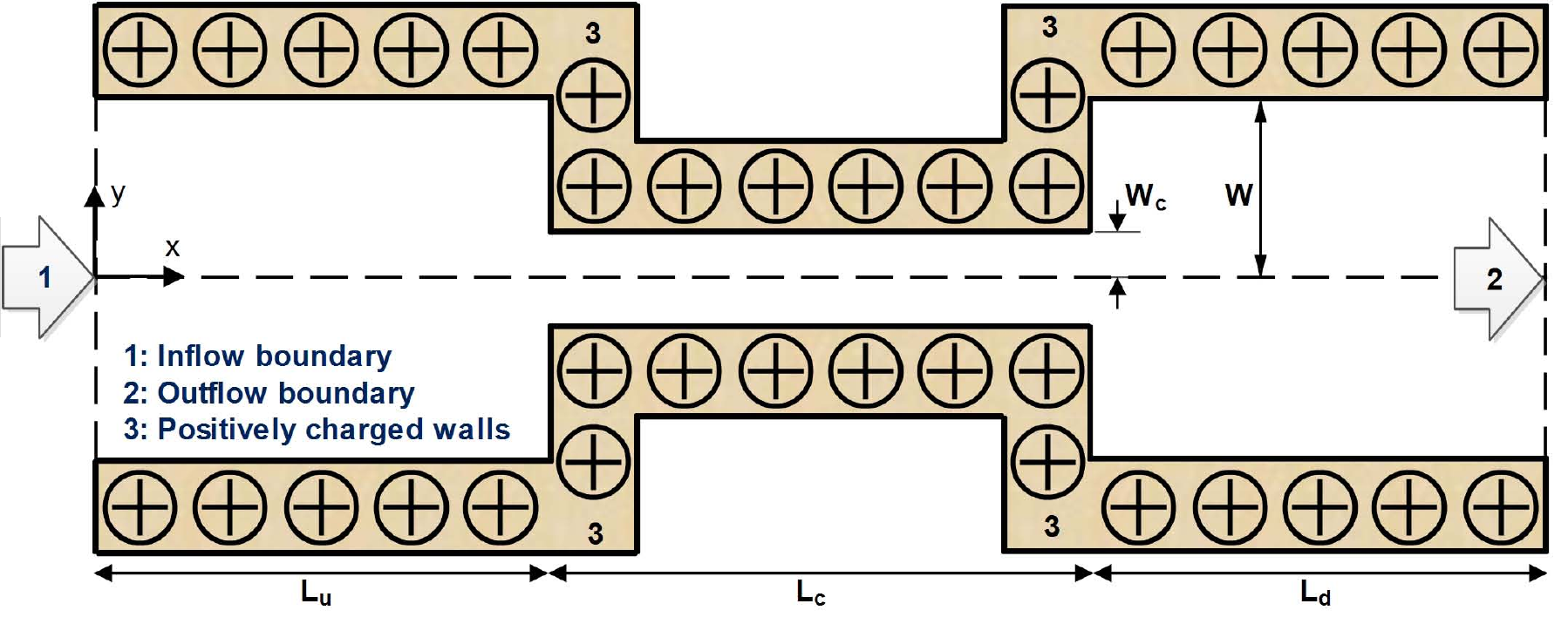}
	\caption{Schematics of electro-viscous flow (EVF) through a contraction-expansion microfluidic device.}
	\label{fig:1}
\end{figure}

\noindent 
Consider the pressure-driven fully developed flow (with an average inflow velocity of $\overline{V}$, m/s) of electrolyte solution through an electrically charged non-uniform (i.e., contraction -- expansion) slit microfluidic device, as shown in \fig\ref{fig:1}. The contraction section is placed in between the upstream inlet and downstream outlet sections. The length (in $\mu$m) of the upstream, downstream, and contraction sections of the device is $L_{\text{u}}$, $L_{\text{d}}$ and $L_{\text{c}}$, respectively. The total length of the microfluidic device is $L = L_{\text{u}}+L_{\text{c}}+L_{\text{d}}$. The cross-sectional width (in $\mu$m) of the upstream, downstream, and contraction sections of the geometry is $2W$, $2W$, and $2W_{\text{c}}$, respectively. The contraction ratio is defined as $d_{\text{c}}=(W_{\text{c}}/W)$.
\\\noindent  
The liquid is assumed to be incompressible and Newtonian, i.e., density ($\rho$, kg/m$^3$), viscosity ($\mu$, Pa.s), and dielectric constant ($\varepsilon_{\text{r}}$) are spatially uniform. The liquid contains symmetric anions and cations with equal valences ($z_{{+}}=-z_{{-}}=z$) and diffusivity of ions ($\mathcal{D}_{{+}}=\mathcal{D}_{{-}}=\mathcal{D}$, m$^2$/s). The bulk \rev{(i.e., geometric mean)} concentration of each ion species is $n_{\text{0}}$ \citep{Harvie2012,Davidson2016}. The surface charge density ($\sigma$, C/m$^2$) is considered uniform over the device walls.  The dielectric constant of the wall is taken to be negligible to that of liquid ($\varepsilon_{\text{r,w}} \lll \varepsilon_{\text{r}}$). 
\subsection{Governing equations}
\noindent
The present physical problem can be mathematically expressed by the theory of electrostatics, conservation of ionic species, momentum, and mass as follows. 

\noindent
According to electrostatics theory, the Poisson’s equation relates the total electrical potential ($U$, V) and the local charge density ($\rho_{\text{e}}$, C/m$^3$) as follows.
\begin{gather} 
	\varepsilon_{\text{0}}\nabla\cdot\varepsilon_{\text{r}}\nabla U=-\rho_{\text{e}}
	\label{eq:A.1}
\end{gather}
where $\varepsilon_{\text{0}}$ and $\varepsilon_{\text{r}}$ are the vacuum permittivity and dielectric constant of the electrolyte liquid, respectively.  

\noindent 
The net charge density for an ideal electrolyte is expressed as
\begin{gather}
	\rho_{\text{e}}=\sum_{j=1}^{N}\rho_{\text{e,j}}
	\qquad\text{where}\qquad \rho_{\text{e,j}} =z_{\text{i}}{e}n_{\text{j}}
	\label{eq:A.2}
\end{gather}
where $n_{\text{j}}$, $z_{\text{j}}$ and $e$ are the number density of $\text{j}^{\text{th}}$ type ion,  chemical valance of $\text{j}^{\text{th}}$ type ion, and elementary charge of a proton, respectively.
\\\noindent 
In the case of the electrokinetic flow, the total potential is typically expressed as the sum of EDL and streaming potentials for uniform cross-section microchannels, i.e., 
\begin{gather} 
	U(x,y)=\psi(y)- xE_{\text{x}}
	\label{eq:A.3}
\end{gather}
where $\psi$, $E_{\text{x}}$ and $x$ are the EDL potential (V), the uniform induced electric field strength (V/m) in the axial direction, and axial distance along the geometry. Since EDL potential is independent of axial direction and streaming potential vary linearly along the channel, the two potentials can be decoupled as the streaming potential field is parallel to the wall of the uniform cross-sectional geometries \citep{bharti2009electroviscous}. It is, however, not possible to split and decouple the two potential fields \citep{davidson2007electroviscous,bharti2008steady,davidson2010electroviscous,VASU20101641,Berry2011}  in the case of the non-uniform cross-sectional geometries like contraction-expansion. 
\\\noindent 
The conservation of each ionic species is expressed by the Nernst-Planck (N-P) equation as follows.  
\begin{gather} 
	\frac{\partial n_{\text{j}}}{\partial t}+\nabla\cdot \myvec{f}_{\text{j}}=0
	\label{eq:A.4}
\end{gather}                                              %
where $\mathbf{f_{\text{j}}}$\rev{,} the flux density of $\text{j}^{\text{th}}$ type ion, \rev{is described by the Einstein relation as follows.} 
\begin{gather}
	%
	\myvec{f}_{\text{j}}=n_{\text{j}}\myvec{V}-\mathcal{D}_{\text{j}}\nabla n_{\text{j}}-\left(\frac{\mathcal{D}_{\text{j}}z_{\text{j}}en_{\text{j}}}{k_{\text{B}}T}\right)\nabla U
	\label{eq:A.5}
\end{gather} 
where $\mathcal{D}_{\text{j}}$, $\myvec{V}$, $k_{\text{B}}$, and $T$ are the diffusivity of $\text{j}^{\text{th}}$ type ion, velocity vector, Boltzmann constant, and temperature, respectively.     
\\\noindent 
The conservation of momentum and mass of an incompressible electrolyte liquid flow can be expressed by the Navier-Stokes (N-S) and mass continuity equations, as follow.
\begin{align}
	\rho\left[\frac{\partial\myvec{V}}{\partial t} + \nabla\cdot (\myvec{V}\myvec{V})\right]
	&= -\nabla P + \nabla\cdot \mu\left[\nabla \mathbf{V} + (\nabla \myvec{V})^T\right] + \myvec{F}_{\text{e}} \label{eq:A.6}	
	\\
	\nabla\cdot\myvec{V}&=0	\label{eq:A.7}
\end{align}
where $t$, $\rho$, $\mu$ and $P$ are the time, density and viscosity of liquid, and pressure, respectively. 
In Cauchy momentum equation (\eqn\ref{eq:A.6}), the extra electrical force due to free charge is given by 
\begin{gather} 
	\myvec{F}_{\text{e}}=-(\rho_{\text{e}}\nabla U)
	\label{eq:A.8}
\end{gather}
The flow field (\eqn\ref{eq:A.6}) is coupled with both electrical potential (Poisson’s equation, \eqn\ref{eq:A.1}) and ion concentration  (Nernst–Planck equation, \eqn\ref{eq:A.4}) fields.
%

%
\noindent 
The governing equations (\eqns\ref{eq:A.1} to \ref{eq:A.8}) are non-dimensionalized by using the following scaling factors: $(k_{\text{B}}T/ze)$, $n_{\text{0}}$, $\overline{V}$, $\rho\overline{V}^2$,$W$, $(W/\overline{V})$ for electrical potential, the number density of ions, velocity, pressure, length, and time, respectively. 
%

\noindent 
The dimensionless form of the governing equations (\eqns\ref{eq:A.1}, \ref{eq:A.4},  \ref{eq:A.6} and \ref{eq:A.7}) is written as follow. The variable names have been retained same as in dimensional equations (\eqns \ref{eq:A.1}-\ref{eq:A.8}) for convenience.
\setlength{\belowdisplayskip}{6pt} \setlength{\belowdisplayshortskip}{6pt}
\setlength{\abovedisplayskip}{6pt} \setlength{\abovedisplayshortskip}{6pt}
\begin{gather}
\nabla^2U=-\rev{\frac{1}{2}K^2}(n_{\text{+}}-n_{\text{-}})
\label{eq:1}
\end{gather}
\begin{gather}
\left[\frac{\partial n_{\text{j}}}{\partial t}+\nabla\cdot(\myvec{V}n_{\text{j}})\right]=\rev{\frac{1}{Pe}}\left[\nabla^2n_{\text{j}}\pm\nabla\cdot(n_{\text{j}}\nabla U)\right]
\label{eq:2}
\end{gather}
\begin{gather}
\left[\frac{\partial \mathbf{V}}{\partial t}+\nabla\cdot(\myvec{V}\myvec{V})\right]=-\nabla P+\rev{\frac{1}{Re}}\nabla \cdot\left[\nabla\myvec{V}+(\nabla\myvec{V})^T\right]-\underbrace{\rev{\beta\left(\frac{K}{Re}\right)^2}(n_{\text{+}}-n_{\text{-}})\nabla U}_{\myvec{F}_{\text{e}}}
\label{eq:3}
\end{gather}
\begin{gather}
\nabla\cdot\myvec{V}=0 \label{eq:4} 
\end{gather}
%
%
\rev{where $U$, $n_{\text{j}}$, $\mathbf{V}$ and $P$ are the total electrical potential, number density of $\text{j}^{\text{th}}$ type ion,  velocity vector, and pressure, respectively.}
The dimensionless groups appearing in \eqns(\ref{eq:1}) to (\ref{eq:4}) are defined as follow.
\begin{gather}
Re=\frac{\rho\bar{V}W}{\mu}, \qquad
\mathit{Sc}=\frac{\mu}{\rho \mathcal{D}}, \qquad
Pe =Re\times\mathit{Sc}, \qquad
\beta=\frac{\rho k_{\text{B}}^2T^2\varepsilon_{\text{0}}\varepsilon_{\text{r}}}{2z^2e^2\mu^2}, \qquad 
K^2=\frac{2W^2z^2e^2n_{\text{0}}}{\varepsilon_{\text{0}}\varepsilon_{\text{r}} k_{\text{B}}T}
\label{eq:5}
\end{gather}
where $Re$, $\mathit{Sc}$, $Pe$, $\beta$, and $K$ are the Reynolds number, Schmidt number, Peclet number, liquid parameter, and inverse Debye length ($K=\lambda_{\text{D}}^{-1}$), respectively.
%
%
\subsection{Boundary conditions}
\noindent The relevant boundary conditions for the mathematical model (\eqns\ref{eq:1} to \ref{eq:4}) specified at the inlet, outlet, and the walls of the microfluidic device are given below. 

\noindent 
(a) At the inlet ($x=0$) of the microfluidic device, velocity and ionic concentration profiles are obtained and imposed from the numerical solution of the steady, fully developed flow of electrolyte liquid through the two-dimensional uniform slit, as follow.
\begin{gather}
	V_{\text{x}}=V_{\text{0}}(y), \qquad
	V_{\text{y}}=0, \qquad
	n_{{+}}=n_{{0}}\exp\left[\frac{-ze\psi(y)}{k_{\text{B}}T}\right], \qquad 
	n_{{-}}=n_{{0}}\exp\left[\frac{ze\psi(y)}{k_{\text{B}}T}\right]
	\label{eq:A.9}
\end{gather}
where $V_{\text{0}}(y)$ and $\psi(y)$ are the fully developed velocity and the EDL potential fields, respectively, for a uniform slit flow. The ionic density  ($n_{{+}}$ and $n_{{-}}$) field is expressed by the Boltzmann equation. 
Analytical and finite-difference (FD) solution procedures to obtain these fields for uniform slit are explained elsewhere  \citep{bharti2008steady,bharti2009electroviscous,davidson2007electroviscous}.
\\
Since the total electrical potential ($U$) appears as a gradient in the field equations (\eqns\ref{eq:A.1} to \ref{eq:A.8}), the axial potential gradient ($\partial U/\partial x$) at the inlet is considered to be uniform. 
In electroviscous flow, the uniform axial potential gradient ($\nabla U$) or the induced electric field ($E=-\nabla U$) is determined such that `zero net current condition' or the `current continuity condition' (\eqn\ref{eq:A.10}) is satisfied.
The net axial induced current density ($I_{\text{net}}=\nabla\cdot I$), i.e., the total current passing across the boundary, becomes zero at a steady-state \citep{bharti2008steady,bharti2009electroviscous,davidson2007electroviscous}. 
The current continuity condition is thus satisfied at the inlet of the microfluidic device, as follows.
\begin{gather}
	\nabla\cdot I = 0\qquad \Rightarrow \qquad 
	I_{\text{net}}=
	\int_{-W}^{W}I_{\text{s}} dy + \int_{-W}^{W} I_{\text{d}}dy + \int_{-W}^{W}I_{\text{c}} dy=0
	\label{eq:A.10}
\end{gather}
where, $I_{\text{s}}$, $I_{\text{d}}$ and $I_{\text{c}}$ are the streaming,  diffusion, and conduction (or faradaic) current densities, respectively, and expressed as follow. 
\begin{gather}
	I_{\text{s}} = \rho_{\text{e}}\mathbf{V}, \qquad
	I_{\text{d}} = -\mathcal{D}\nabla \rho_{\text{e}};
	\qquad\text{and}\qquad
	I_{\text{c}} =-{\sigma_{\text{e}}\nabla U}
	\label{eq:A.11}
\end{gather}
where, the electrical conductivity ($\sigma_{\text{e}}$) of an electrolyte solution, i.e., a net contribution from all ions, is expressed as follows.
\begin{gather} 
	\sigma_{\text{e}} = \sum_{j=1}^{N} \left(\frac{\mathcal{D}_{\text{j}}z_{\text{j}}e}{k_{\text{B}}T}\right)\rho_{\text{e,j}} 
	\label{eq:A.12}
\end{gather}
Further, the diffusion current becomes zero ($I_{\text{d}}=0$) at the steady state condition.
In \eqn(\ref{eq:A.10}), all quantities are calculate at the inlet ($x=0$) of the device. 
\\
(b) At the outlet ($x=L$) of the device, the velocity and ion concentration fields are allowed to be fully developed, i.e.,
\begin{gather}
	\frac{\partial \mathbf{V}}{\partial \myvec{n}_{\text{b}}} = 0,\qquad \text{and}\qquad
	\frac{\partial n_{\text{j}}}{\partial \myvec{n}_{\text{b}}} = 0
	\label{eq:A.13}
\end{gather}
where $\myvec{n}_{\text{b}}$ is outward unit vector normal to the boundary. 
\\\noindent The uniform axial potential gradient is also imposed at the outlet by satisfying the net axial current condition ($I_{\text{net}} = 0$, \eqn\ref{eq:A.10}) in conjunction with zero diffusion current ($I_{\text{d}}=0$) at the steady-state.  To satisfy the current continuity condition on the outlet, all quantities of \eqn(\ref{eq:A.10}) are calculate at the outlet ($x=L$) of the device.
\\
(c) On the device walls, a zero flux density of ions, normal to the uniformly charged solid impermeable wall boundaries ($V_{\myvec{n}_{\text{b}}} =0$), is imposed as follows. 
%
\begin{gather} 
	\mathbf{f}_{\text{j}}\cdot \myvec{n}_{\text{b}}=0,
	\label{eq:A.14}
\end{gather}
Uniform surface charge density is assumed at the walls of microfluidic device. It is expressed as follows.
\begin{gather} 
	\varepsilon_{\text{0}}\varepsilon_{\text{r}}(\nabla U\cdot\myvec{n}_{\text{b}}) = \sigma
	\label{eq:A.14a}
\end{gather}
where $\sigma$ denotes the uniform surface charge density at the device walls.
\\\noindent 
Further, the wall velocity is imposed as a surface charge-dependent slip velocity condition \citep{jing2015coupling} and expressed as follows.
\begin{gather}
	V_{\myvec{t}_{\text{b}}}=b\frac{\partial \myvec{V}}{\partial \myvec{n}_{\text{b}}}
	\qquad\mbox{and}\qquad 
	V_{\myvec{n}_{\text{b}}} =0
	\label{eq:A.15}
\end{gather}
where $V_{\myvec{t}_{\text{b}}}$, and $V_{\myvec{n}_{\text{b}}}$ are the tangential and normal components of the wall velocity. Since the surface charge density ($\sigma$) can affect the slip length \citep{Yang42003,Tian122021} in the microfluidic flow, this effect should be considered during the analysis of the electroviscous flow.  The surface charge-dependent slip ($b$) length is expressed \citep{jing2015coupling,joly2006liquid} as follow.
\begin{gather} 
	b=\frac{b_{\text{0}}}{1+\sigma^2X_{\sigma}b_{\text{0}}}
	\label{eq:A.16} \\
	\text{where}\qquad 
	X_{\sigma} = \frac{1}{\alpha}\left(\frac{d^2l_{\text{B}}}{e^2}\right),~\text{nm}^{3}\text{C}^{-2}\qquad \text{and}\qquad
	l_{\text{B}}=\frac{e^2}{4\pi\varepsilon_{0}\varepsilon_{\text{r}}k_{\text{B}}T},~\text{nm}
	\nonumber
\end{gather}
where $b_{0}$, $\alpha$ ($\sim 1$), $d$ (= 0.4 nm), and $l_{\text{B}}$ are slip length in absence of surface charge, numerical factor, equilibrium distance of Lennard-Jones potential, and Bjerrum length, respectively.

\noindent
The dimensionless form of the boundary conditions (\eqns\ref{eq:A.9} to \ref{eq:A.16}) is expressed as follows.
\noindent 
\rev{(a) At the inlet ($x=0$) of the microfluidic device, the dimensionless form of the conditions (\eqn\ref{eq:A.9}) is expressed as follows.}
\begin{gather}
V_{\text{x}}=V_{\text{0}}(y), \qquad
V_{\text{y}}=0, \qquad
n_{{+}}=\exp[{-\psi(y)}], \qquad 
n_{{-}}=\exp[{+\psi(y)}]
\label{eq:6}
\end{gather}
%
%
\rev{(b) At both inlet ($x=0$) and outlet ($x=L$), the dimensionless form of the `current continuity condition' (\eqn\ref{eq:A.10}) is expressed as follows.}
%
\begin{gather}
\nabla\cdot I = 0\qquad \Rightarrow\qquad I_{\text{net}} = 
\int_{-1}^{1}I_{\text{s}} dy + \int_{-1}^{1} I_{\text{d}}dy + \int_{-1}^{1}I_{\text{c}} dy=0 
\label{eq:7}
\end{gather}
\rev{where, }
\begin{gather}
I_{\text{s}} = {(n_{\text{+}}-n_{\text{-}})\myvec{V}}, \qquad
I_{\text{d}} = -{\rev{\frac{1}{Pe}}\left[\frac{\partial n_{\text{+}}}{\partial x}-\frac{\partial n_{\text{-}}}{\partial x}\right]}, 
\qquad
I_{\text{c}} =-{\rev{\frac{1}{Pe}}\left[(n_{\text{+}}+n_{\text{-}})\frac{\partial U}{\partial x}\right]}\quad 
\label{eq:8}
%
\end{gather}
%
\rev{(c) At the outlet ($x=L$) of the devicet, the dimensionless form of the condition (\eqn\ref{eq:A.13}) is expressed as follows.}
\begin{gather}
\frac{\partial \myvec{V}}{\partial \myvec{n}_{\text{b}}} = 0,
\qquad
\frac{\partial n_{\text{j}}}{\partial \myvec{n}_{\text{b}}} = 0, \qquad P = 0
\label{eq:9}
\end{gather}
\rev{(d) At the solid walls, the dimensionless form of the condition (\eqn\ref{eq:A.14}) is expressed as follows.}
\begin{gather}
\myvec{f}_{\text{j}}\cdot \myvec{n}_{\text{b}}=0,
\label{eq:9a}
\end{gather}
\rev{\eqn(\ref{eq:A.15})in the dimensionless form is written as follows.}
\begin{gather} 
\nabla U\cdot\myvec{n}_{b} = S,\qquad\text{where}\qquad S = \frac{ze\sigma W}{\varepsilon_{\text{0}}\varepsilon_{\text{r}} k_{\text{B}}T}
\label{eq:10}
\end{gather}
\rev{where $S$ is the dimensionless surface charge density.}
\noindent
\rev{\eqn(\ref{eq:A.16}) is expressed in the dimensionless form as follows.}
\begin{gather}
V_{\myvec{n}_{\text{b}}} =0,
\qquad 
V_{\myvec{t}_{\text{b}}}=B\frac{\partial \myvec{V}}{\partial \myvec{n}_{\text{b}}}
\label{eq:11}
\end{gather}
where
\begin{gather}
B =  \frac{B_0}{1+(S^2X_{S})B_0},
\qquad
X_{S} = \frac{1}{W}\left(\frac{e}{4\pi z l_{\text{B}}}\right)^2X_{\sigma} 
\label{eq:11a}
\end{gather}
where $B$ is the dimensionless surface charge-dependent slip length\rev{, and $B_0$ is the dimensionless surface charge-independent slip length}. 

\noindent 
The above detailed mathematical model (i.e., coupled governing partial differential equations, based on Poisson's, N-P, and N-S equations, subject to the boundary conditions) is solved numerically by using the finite element method (FEM) to obtain the flow ($\myvec{V}, P$), electrical potential ($U$) and charge concentration ($n_{\pm}$) fields.     
These numerical fields are post-processed to obtain the excess charge distribution ($n^{\ast}=n_{+}-n_{-}$), axial induced electrical field strength ($E_{\text{x}}$), pressure drop ($\Delta P$) and electroviscous correction factor ($Y$). 
%
\section{Numerical approach}
\label{sec:sanp}
%
\noindent 
In this work, the finite element method (FEM)  based computational fluid dynamics (CFD) solver COMSOL multiphysics has been used to obtain the numerical solution of a mathematical model governing the electrolyte liquid flow through the contraction-expansion microfluidic device. The two-dimensional (2-D) fully-coupled multiphysics has been represented by \textit{electrostatics} (es), \textit{transport of dilute species} (tds), and \textit{laminar flow} (spf) modules of COMSOL. The computational domain has been discretized by the linear shape function, uniform (except boundary and corner refinements), rectangular, structured mesh structure. The partial derivatives and partial differential equations (PDEs) have been transformed to the simultaneous algebraic equations (SAEs) by using the finite element method (FEM). The polynomials of first order (P$_\text{p}$+P$_\text{q}$ with $\text{p}=\text{q}=1$), i.e., shape function with first order element, are used for the spatial discretization of the velocity and pressure fields. The integral in \eqn(\ref{eq:7}) is evaluated using the $intop$ function in the model coupling defined in the global function definition. Further, the set of SAEs has iteratively been solved using a fully coupled  PARDISO (PARallel DIrect SOlver) and Newton's non-linear solvers. The steady-state solution yields the total electrical potential ($U$), induced electrical field ($E_{\text{x}}$), pressure ($P$), velocity ($\myvec{V}$), and the ion concentration ($n_{\pm}$) fields.

\noindent
In the present work, the following geometrical (\fig\ref{fig:1}) parameters are considered for the physical system: $L_{\text{u}}=L_{\text{c}}=L_{\text{d}}=5W$, and $d_{\text{c}}=W_{\text{c}}/W=0.25$. Based on the previous knowledge \citep{davidson2007electroviscous,bharti2008steady} and present domain independence test (results not shown here), the lengths of individual sections of the device are experienced to be sufficiently large to ignore the all effects due to entry\rev{, and} exit. Further, the mesh independence tests are performed, in this work, with M1 = 50, M2 =100, and M3 = 150 grid points distributed uniformly per unit length/width of boundaries of the microchannel.  The corner refinement and boundary layer are also considered in all meshes. The results of total electrical potential, charge, induced electric field, and pressure drop have shown insignificant (i.e., $\pm 1-2\%$) variation with the grid refinement from M2 to M3 over the ranges of conditions. Thus, the mesh M2 consisting of 333600 elements (degree of freedom DoF = 3018814) is believed to be sufficiently refined to obtain the final accurate results, free from mesh and end effects.
\section{Results and discussion}
\noindent
This section presents and discusses the numerical results for symmetric (1:1) electrolyte liquid flow through a contraction-expansion (4:1:4) slit microfluidic device for the following ranges of conditions:  Reynolds number ($Re=10^{-2}$), Schmidt number ($\mathit{Sc}=10^{3}$, based on water properties at 298 K), liquid parameter ($\beta=2.34\times10^{-4}$), inverse Debye length ($K=2$, 4, 6, 8, and 20), surface charge density ($S=$0, 4, 8 and 16), and slip length ($B_{0}=0$, 0.05, 0.10, 0.15 and 0.20). Further, $S=0$ (or $K=\infty$) corresponds to the non-electroviscous flows.
\\\noindent
In particular, Reynolds number is taken to be low ($Re=0.01$) as the microfluidic flows are generally steady and laminar \citep{davidson2007electroviscous}. The variation of EDL thickness is accounted  by the inverse Debye length ($K = 2$ for thick EDL, i.e., tending to overlap in channel, and  $K=20$ for very thin EDL). The surface charge density ($S$) accounts for the practical ranges of zeta potential variation from $50$ to $100$ mV when $K=2$ (overlapping EDL), and from 12 to 50 mV when $K=8$ for a uniform microchannel with the variation of $S$ from 4 to 16. \rev{The dimensional surface charge density ($\sigma$) is thus considered in the range of $7.28\times10^{-4}$ (at $S=4$) to $2.91\times10^{-3}$ C/m$^2$ (at $S=16$)} \citep{davidson2007electroviscous}. The slip length ($B_0$) is taken in the range of $0$ to $0.20$ because the lowest $B_0$ ($=0$) express the no-slip condition, and the highest $B_0$ ($=0.20$) shows the higher slippery surface of the microchannel \citep{buren2018electroviscous}.
\noindent
Before presenting new results, a thorough validation of the numerical approach has been performed with the existing relevant literature \citep{davidson2007electroviscous} for limiting no-slip flow condition. The detailed comparisons, however, are not shown here independently to avoid the repetition, but presented in the results section. Both present and literature values have shown excellent ($\pm 1-2\%$) agreement to each other for all field variables ($U$, $n^{\ast}$ and $P$). \rev{However, none of the results are available in the literature for the slip flow in the considered geometry.} The results presented hereafter are, thus, reliable with an excellent ($\pm 1-2\%$) level of accuracy. Subsequently, the new results, based on total 80 simulations performed, are presented for the dimensionless total electrical potential ($U$),  dimensionless excess charge ($n^{\ast}=n_{+}-n_{-}$),  dimensionless pressure ($P$),  dimensionless induced electric field ($E_{\text{x}}$), and the electroviscous correction factor ($Y$) as a function of dimensionless parameters ($K$, $S$ and $B_0$).
%
\subsection{Total electrical potential ($U$) distribution}
\begin{figure}[t]
	\centering\includegraphics[width=1\linewidth]{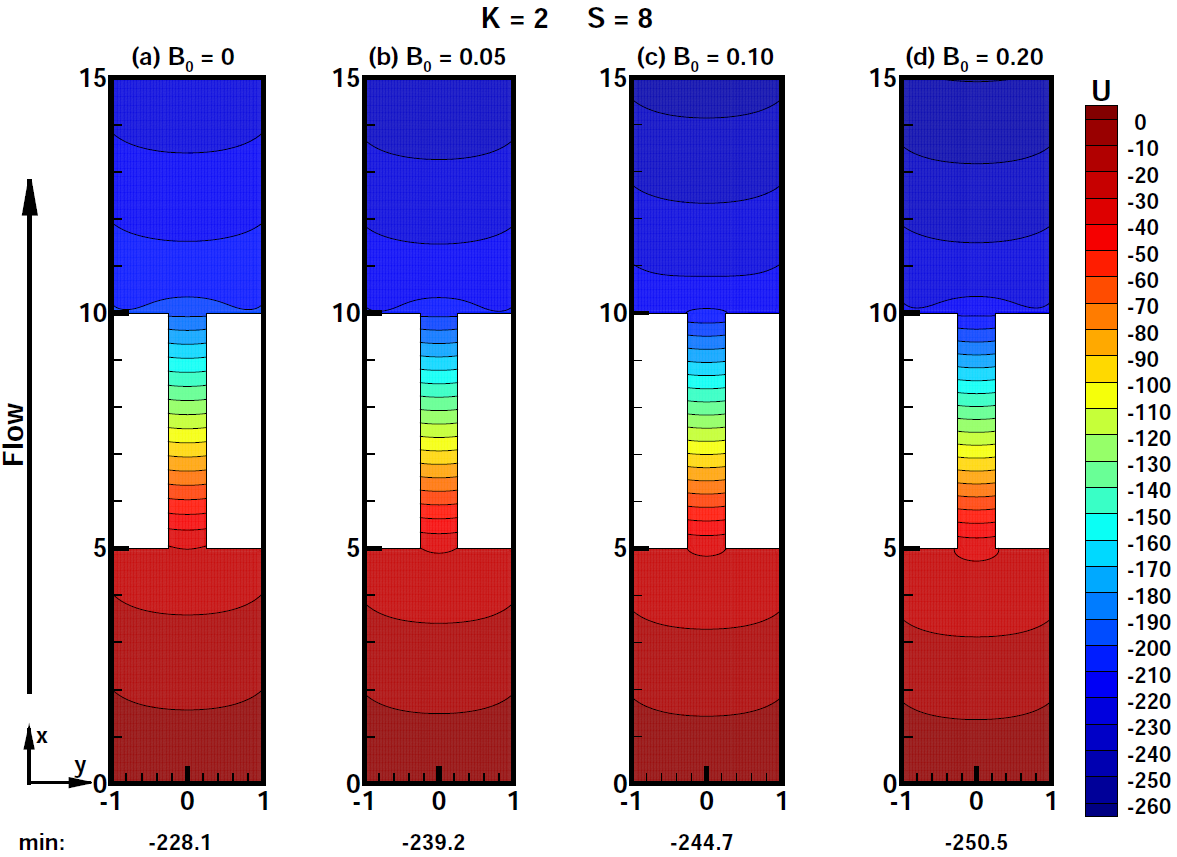}
	\caption{Total electrical potential ($U$) distribution for $B_{\text{0}}=0$ to $0.20$ at $S=8$ and $K=2$.}
	\label{fig:5}
\end{figure}

%
\noindent
\fig\ref{fig:5} shows the distribution of the dimensionless total electrical potential ($U$) in the microfluidic device for the wide range of slip length ($0\le B_{\text{0}}\le 0.2$) at the fixed values of $K=2$ and $S=8$. 
The total electrical potential contours have shown qualitatively similar variations over the ranges of conditions ($K\le 20$, $4\le S\le 16$, and $0\le B_{\text{0}}\le 0.2$) explored herein.  Broadly, the total electrical potential decreases along the length of the device, irrespective of the values of the governing parameters ($K$, $S$ and $B_0$). It is because of the enhancement of negatively charged ions (i.e., excess charge) due to a positively charged surface, which increases the streaming current and decreases the streaming potential. The decreasing streaming potential reduces total electrical potential, as EDL potential remains invariant along the length of the channel,  \rev{except near both ends of the contraction section}. The lateral curving of the electrical potential contours is obtained as the normal potential gradient at the wall is considered equal to surface charge density (\eqn\ref{eq:10}).
The profiles observed here are well consistent with the existing literature \citep{davidson2007electroviscous,davidson2008electroviscous,bharti2008steady,bharti2009electroviscous}.
The contours for other conditions are not shown here due to their qualitatively similar nature. 
\begin{figure}[h]
	\centering\includegraphics[width=1\linewidth]{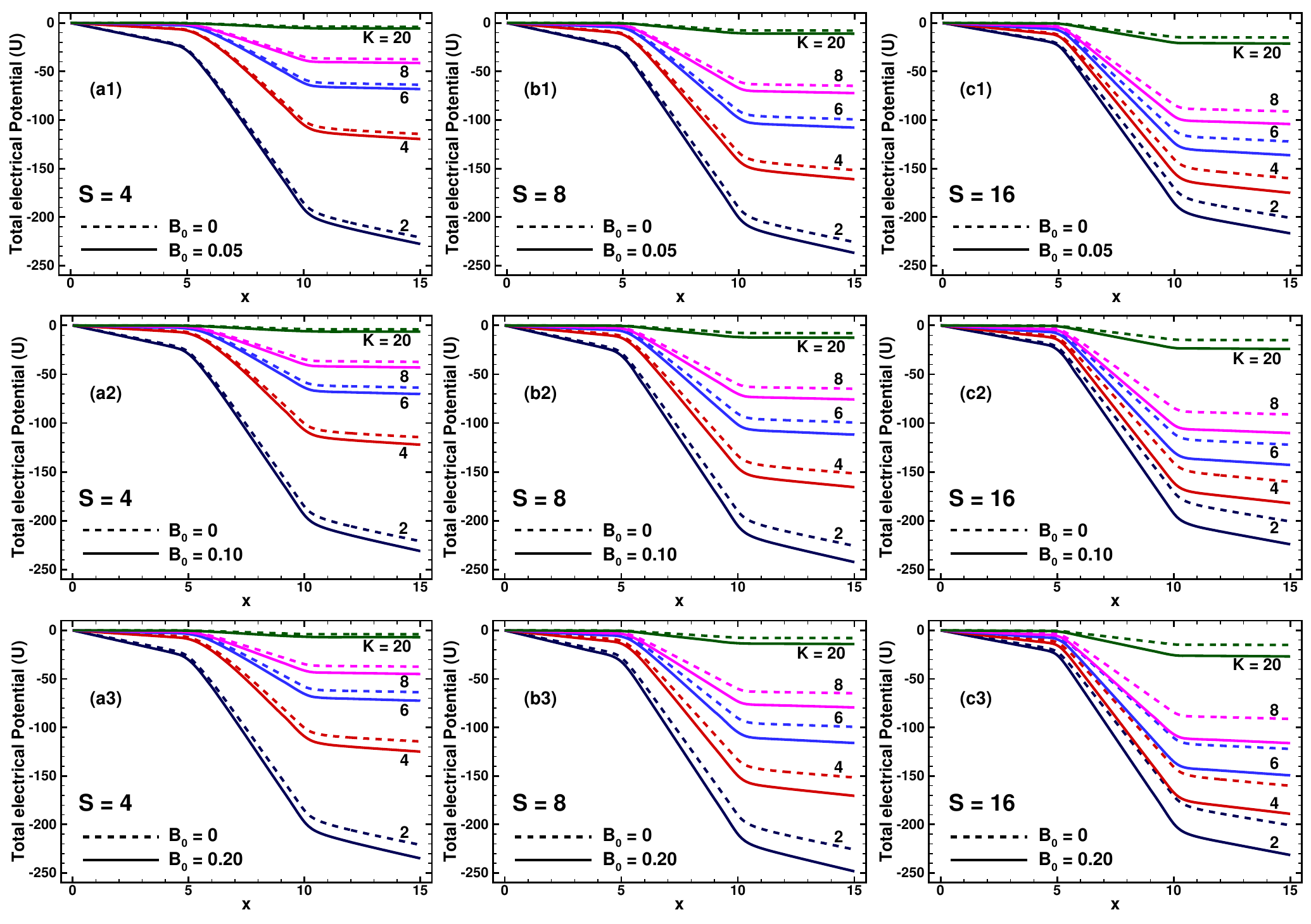}
	\caption{Axial variation of dimensionless total electrical potential ($U$)  along the horizontal centreline ($x,0$)  of the microfluidic device as a function of dimensionless parameters ($K$, $S$ and $B_0$).}
	\label{fig:6}
\end{figure} 

\noindent 
Further, \fig\ref{fig:6} depicts the axial variation of total electrical potential  along the horizontal centreline ($x,0$)  of the microfluidic device for the explored ranges of conditions ($2\le K\le 20$, $4\le S\le 16$, and $0\le B_{\text{0}}\le 0.2$). 
In the flow direction, total potential \rev{drop} ($\Delta U$) decreases due to the advection of fluid having excess negative ions along the length of positively charged microfluidic device. The potential gradient is maximum in the contraction section due to increased convective velocity with the reduction in flow area. The total potential decreases with increasing the slip length ($B_0 > 0$) in comparison to that of no-slip ($B_0=0$) case because the convection velocity near the slip wall increases with increasing slip length. It enhances the excess charge transport and increases the streaming current but decreases the streaming potential. Thus, total electrical potential decreases with increasing slip length. The reduction in the potential is less at lower $S$ (as shown in \fig\ref{fig:6}a) and more at higher $S$ (as shown in \fig\ref{fig:6}c).
\begin{table}
\centering\caption{The total electrical potential drop ($\Delta U$) along the horizontal centreline ($x,0$) over the length of the microfluidic device.}\label{tab:1}
\begin{tabular}{|r|r|r|r|r|r|r|}
\hline
$S$	&	$K$	&	\multicolumn{5}{c|}{$\Delta U$}	\\\cline{3-7}
	&		&	$B_0=0$	&	$B_0=0.05$	&	$B_0=0.10$	&	$B_0=0.15$	&	$B_0=0.20$	\\\hline
0	&	$\infty$	&	0	&0		&0		&0		&0		\\\hline
4	& 2	& -220.9700	& -227.6700	& -231.0700	& -233.2800	& -234.8800 \\
& 4	& -114.4300	& -119.6000	& -122.1500	& -123.7800	& -124.9300 \\
& 6	& -63.6860	& -68.1300	& -70.26700	& -71.5830	& -72.4950 \\
& 8	& -37.4060	& -41.2680	& -43.1050	& -44.2180	& -44.9780 \\
& 20	& -3.9929	& -5.6068 	& -6.3591 	& -6.8023	& -7.0975 \\\hline
8	& 2	& -225.7700	& -236.9100	& -242.4300	& -245.8600	& -248.2500 \\
& 4	& -151.5500	& -161.0100	& -165.6300	& -168.4900	& -170.4700 \\
& 6	& -99.3690	& -107.8300	& -111.8800	& -114.3500	& -116.0400 \\
& 8	& -64.8040	& -72.2580	& -75.7920	& -77.9210	& -79.3660 \\
& 20	& -7.8852	& -11.0860	& -12.5750	& -13.4510	& -14.0340 \\\hline
16	& 2	& -200.7800	& -216.7400	& -224.1600	& -228.5300	& -231.4300 \\
& 4	& -160.1100	& -175.0800	& -182.0700	& -186.2100	& -188.9700 \\
& 6	& -122.1500	& -136.2600	& -142.8700	& -146.7900	& -149.4100 \\
& 8	& -91.1770	& -104.1500	& -110.2000	& -113.7800	& -116.1800 \\
& 20	& -15.0310	& -21.2220	& -24.0770	& -25.7490	& -26.8570 \\\hline
\end{tabular}
\end{table}

\noindent 
\tab\ref{tab:1} summarizes the total electrical potential drop ($\Delta U$) along the horizontal centreline ($x,0$)  over the length of the microfluidic device for the ranges of explored conditions.
Quantitatively, the potential drop increases with increasing $K$, and a minimal reduction is observed at $K=20$ for all values of $S$ and $B_0$. Further, \rev{the magnitude of} $\Delta U$ increases with increasing $S$ (as shown in \fig\ref{fig:6} and \tab\ref{tab:1}), except at $S=16$ and $K=2$. \rev{
It is because the effective excess charge available for the transport in the EDL  decreases, thereby decreasing the streaming current and the total potential drop ($\Delta U$) with increasing $S$ at higher $K$ where the EDLs are not overlapping. The trends, however, reverse at lower $K$ where EDLs tend to overlap, resulting in an enhancement in available excess charge for transport and magnitude of $\Delta U$ increases.}
Thus, \rev{$|\Delta U|$} increases with increasing $S$, except at higher $S(=16)$ with overlapping EDL ($K=2$). Further, $\Delta U$ decreases with increasing $B_0$ for a given value of $K$.   
For instance, the \rev{magnitude of} $\Delta U$ increases by 98.19\%, 96.51\% and 92.51\% at $S=4$, 8 and 16, respectively, with decrease in $K$ from 20 to 2 for no-slip ($B_0=0$) condition. The corresponding drops in $|\Delta U|$ are noted as 97.54\%, 95.32\% and 90.21\% for increase in slip length from $B_0=0$ to 5\%. The values of $\Delta U$, however, drops by 96.98\%, 94.35\% and 88.4\% with increases in $B_0=0$ to 20\%. 
As $K$ increases, EDL thickness reduces, and the electrical potential distributes in the close vicinity of the wall, i.e., sharp potential gradient ($\partial U/\partial n$) normal to the charged wall. This redistribution of electrical potential in the close vicinity of the wall, in turn, reduces the axial potential drop ($\partial U/\partial x$). \rev{It is because} of reduction in the available free charge (negative ions) in the EDL for transport that decreases the streaming current and hence streaming potential with increasing K (or thinning of EDL). 
Furthermore, the relative drop\footnote{relative change in quantity $\phi$ for a change in any variable from $p$ to $q$ defined as $\phi_r = (\phi_{q} - \phi_{p})/\phi_{p}$} in the electrical potential ($\Delta U_r$) increases with increasing slip length ($B_0$), irrespective of the values of $S$ and $K$. Further, the electrical potential drop also increases with increasing $K$ and decreases with increasing $S$, irrespective of $B_0$.
For instance, the potential drop ($\Delta U_r$) increases by 3.03\% (at $K=2$) and 40.42\% (at $K=20$) with increase in the slip length ($B_0$) from 0 to 5\% for $S=4$. The corresponding drop in potential increases by 7.95\% (at $K=2$) and 41.19\%  (at $K=20$) for $S=16$.
Similarly, $\Delta U_r$ increases by 6.29\% (at $K=2$) and 77.75\% (at $K=20$) with increase in the slip length ($B_0$) from 0 to 20\% for $S=4$. The corresponding $\Delta U_r$ increases by 15.27\% (at $K=2$) and 78.68\%  (at $K=20$) for $S=16$.

\noindent The functional dependence of the total potential drop ($\Delta U$, \tab\ref{tab:1}) along the horizontal centreline ($x,0$)  over the length of the microfluidic device on the dimensionless governing parameters ($K$, $S$ and $B_0$) can be expressed by the following predictive correlation.
\begin{gather}
\Delta U = \sum_{i=1}^{4} A_{\text{i}} (\ln K)^{(i-1)}
\label{eq:12}
\\
\qquad\text{where}\quad A_{\text{i}} = \sum_{j=1}^{3} A_{\text{ij}} B_0^{n},\qquad 
A_{\text{ij}} = \sum_{k=1}^{3} M_{\text{ijk}}S^{(k-1)} 
\qquad\text{and}\quad n=\frac{(j-1)(6-j)}{4} \nonumber
\end{gather}
The correlation coefficients ($M_{\text{ijk}}$) are statistically obtained, for 75 data points, as by performing the non-linear regression analysis using the DataFit (trial version).
\begin{gather*}
M = \begin{bmatrix} M_1 & M_2 & M_3 & M_4\end{bmatrix}^T
\\
M_1 = \begin{bmatrix}
499.51	&	-43.43	&	1.445	\\
9.7831	&	52.128	&	-1.2754	\\
32.664	&	-82.825	&	2.0086	
\end{bmatrix}, 
\quad 
M_2 = \begin{bmatrix}
-551.09	&	107.67	&	-4.127	\\
19.708	&	-24.538	&	0.6022	\\
-100.91	&	55.55	&	-1.6347		
\end{bmatrix}, 
\end{gather*}
\begin{gather*}
M_3 = \begin{bmatrix}
197.16	&	-58.106	&	2.4119	\\
-26.892	&	13.921	&	-0.1944	\\
73.756	&	-32.705	&	0.799		
\end{bmatrix}, 
\quad
M_4 = \begin{bmatrix}
-22.989	&	9.0533	&	-0.3992	\\
6.3935	&	-3.4275	&	0.0437	\\
-14.565	&	7.1861	&	-0.1573	
\end{bmatrix} 
\end{gather*}
with $\delta_{\text{min}}=-3.28\%$, $\delta_{\text{max}}=1.48\%$, $\delta_{\text{avg}}=-0.98\%$ and $R^2=99.99\%$ for the range of the conditions explored herein. Here, $\delta_{\text{min}}$, $\delta_{\text{max}}$, $\delta_{\text{avg}}$ and $R^2$ being the minimum, maximum and average deviation from the numerical values and coefficient of determination, respectively.

\noindent 
The total electrical potential distribution in the microfluidic devices resulted from the complex interplay of the dimensionless parameters ($K$, $S$, and $B_0$). The EDL (electrical double layer) thickness decreases with increasing $K$, and it increases with increasing $S$. An increasing slip length ($B_0>0$) further assists the advection of ions with the flow along the length of the microfluidic device. As the Poisson equations (\eqn\ref{eq:1}) relate the distribution of electrical potential with charge, the subsequent section explores the distribution of the excess charge ($n^\ast$) as a function of the flow governing parameters.  
%
\subsection{Excess charge ($n^\ast$) distribution}
%
\noindent 
The difference between the positive ($n_{+}$) and negative ($n_{-}$) ion concentrations is denoted as the  excess ionic number concentration ($n^{\ast}=n_{+}-n_{-}$). It is also referred to as excess charge ($n^\ast$), as it equals the dimensionless net charge density ($\rho_{\text{e}}$) for the symmetric electrolyte solution.
\begin{figure}[htbp]
	\centering\includegraphics[width=0.95\linewidth]{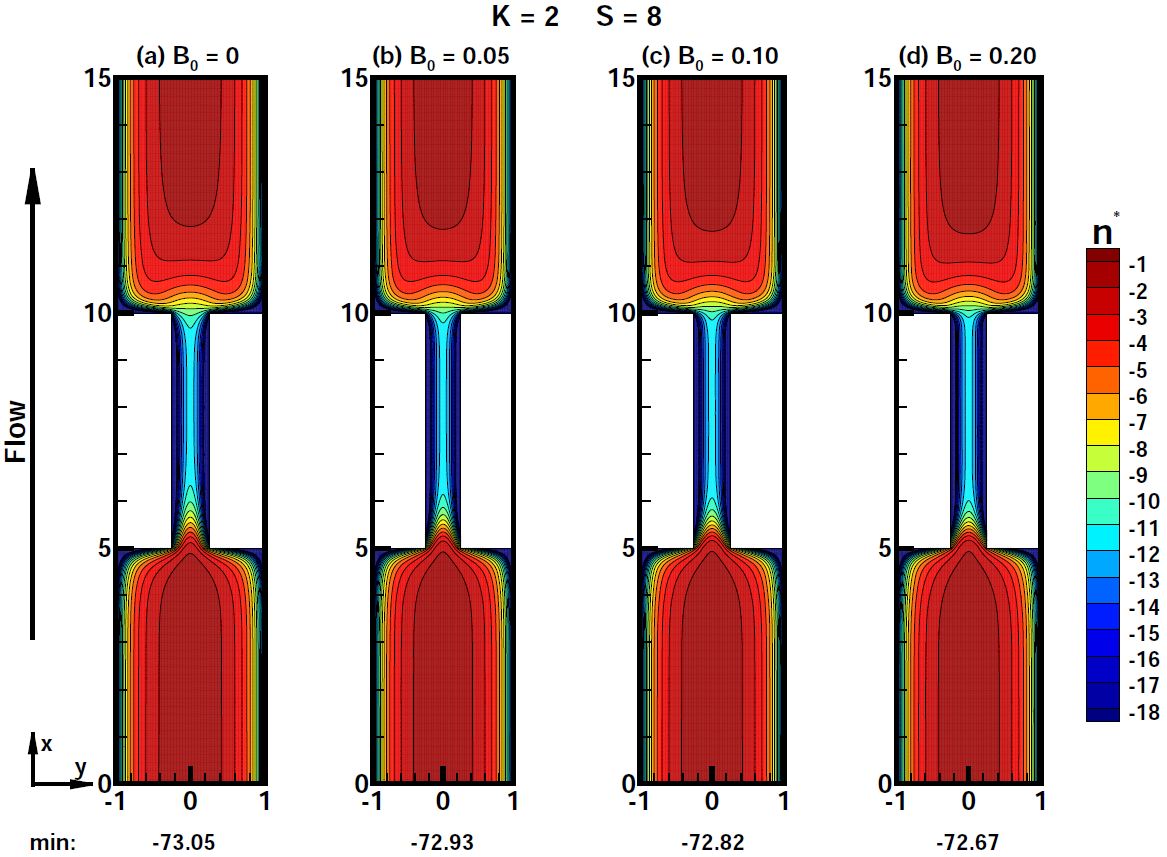}
	\caption{Dimensionless charge ($n^\ast$) distribution profiles  as a function of $B_{\text{0}}$ at $S=8$ and $K=2$.}
	\label{fig:7}
\end{figure} 
\fig\ref{fig:7} depicts the excess charge distribution in the microfluidic device for various slip lengths ($B_{\text{0}}$) at fixed values of $K=2$ and $S=8$. Qualitatively similar profiles are observed for other conditions and thus not presented here. 
The charge distribution, in general, has shown complex dependence on the flow governing parameters ($K$, $S$ and $B_0$).  
For instance, the excess charge is obtained negative ($n^\ast < 0$) throughout the device for all conditions. It suggests the prominence of the negative ions ($n_{-}$) for the positively charged ($S>0$) surface. The high-density clustering of the excess charge is noticeable in the close vicinity of the charged walls, irrespective of the flow conditions. The clustering of contours is further dense in the contraction section due to suddenly converging flow area. It is attributed to the attraction of the negative ions and repulsion of the positive ions of the electrolyte solution by the positively charged surface and vice versa for the negatively charged surface. 
\begin{figure}[h]
	\centering\includegraphics[width=1\linewidth]{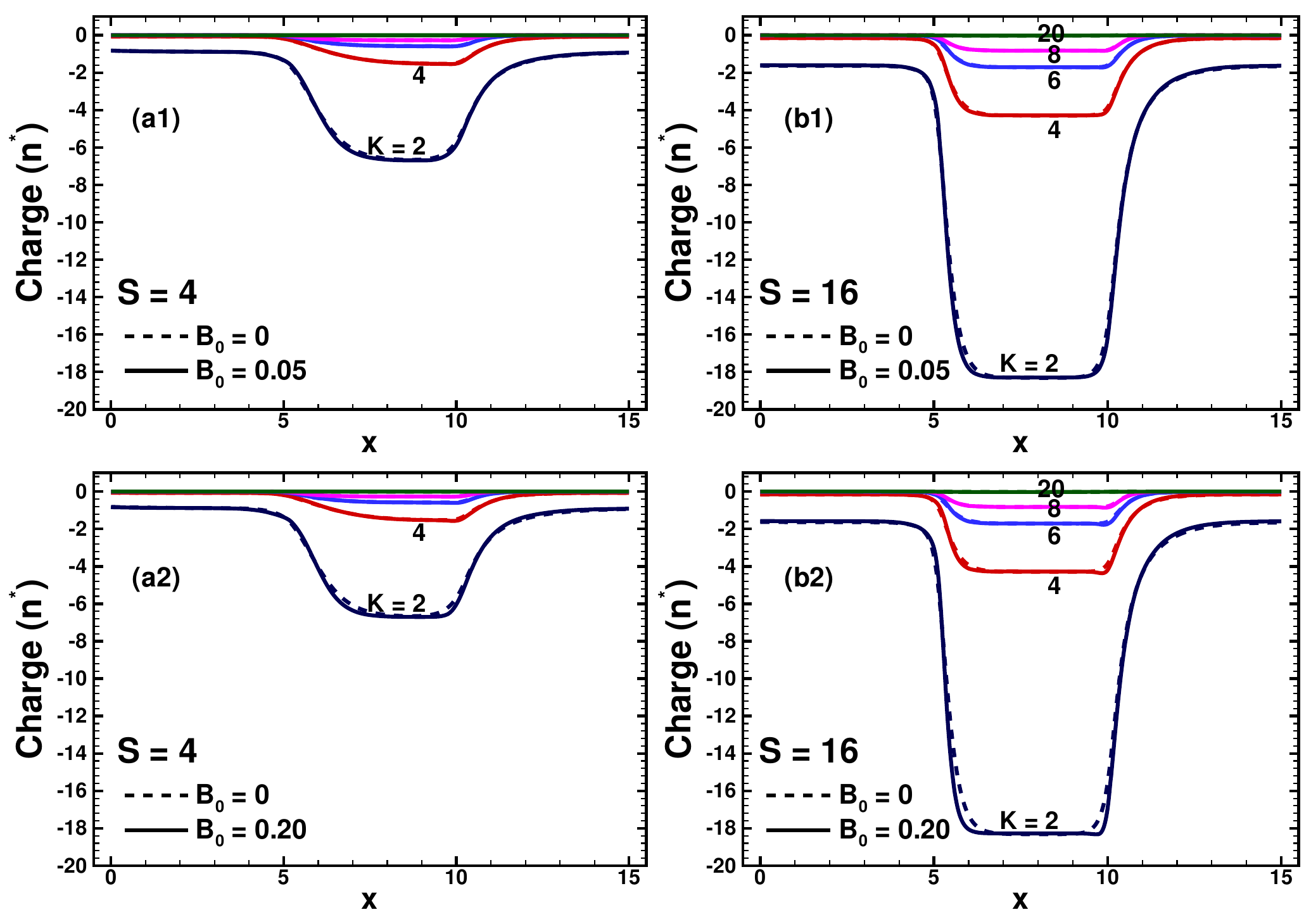}
	\caption{Axial variation of dimensionless excess charge ($n^\ast$) along the horizontal centreline ($x,0$)  of the microfluidic device as a function of dimensionless parameters ($K$, $S$ and $B_0$).}
	\label{fig:7v}
\end{figure} 
Furthermore, the minimum value of excess charge in the device has shown negligible influence due to slip ($B_0$) intensity. For instance, the magnitude of the minimum excess charge ($|n^\ast_{\text{min}}|$) is noted invariable with $B_0$ as -73.05, -72.93, -72.82 and -72.67 for $B_0=0$, 5\%, 10\% and 20\%, respectively, at $K=2$ and $S=8$. Similarly, $|n^\ast_{\text{min}}|$ values are recorded as ($\sim$19, $\sim$73 and $\sim$292) and ($\sim$1, $\sim$2 and $\sim$4) for ($S=4$, 8 and 16), irrespective of $B_0$, at $K=2$ and 20, respectively.

\noindent 
\fig\ref{fig:7v} shows the axial variation of excess charge over the horizontal centreline ($x,0$)  of the microfluidic system for the ranges of governing parameters ($K$, $S$ and $B_0$). As evident through contour profiles (\fig\ref{fig:7}), the excess charge is equal and most prominent (\fig\ref{fig:7v}) at the centreline locations of both inlet and outlet of the device. The excess charge decreases along the device length from the inlet/outlet to the contraction section. The magnitude of the excess charge ($|n^\ast|$) is highest in contraction than upstream/downstream sections. The trends remain same for all values of $K$, $S$ and $B_0$. The results reflect the stronger dependency of $n^\ast$ on both $K$ and $S$ 
in comparison to that on $B_0$.  Overall, the charge distribution shows complex dependence on $K$, $S$ and $B_0$. It is due to the ionic species transfer being highly dependent on the charge capacity of the walls than the convection velocity near the surface.  The charge distribution behaviours shown in \figs\ref{fig:7} and \ref{fig:7v} are consistent with the existing literature \citep{davidson2007electroviscous} for the no-slip ($B_0=0$) condition.

\noindent 
Further, \tab\ref{tab:2} comprises the minimum values of the excess charge ($n^\ast_{\text{min}}$) over the horizontal centreline ($x,0$)  of the microfluidic device as a function of dimensionless parameters ($K$, $S$ and $B_0$). 
\begin{table}
\centering\caption{Minimum values of excess charge ($n^\ast_{\text{min}}$) over the horizontal centreline ($x,0$)  of the microfluidic device.}\label{tab:2}
\begin{tabular}{|r|r|r|r|r|r|r|}
\hline
$S$	&	$K$	&	\multicolumn{5}{c|}{$n^\ast_{\text{min}}$}	\\\cline{3-7}
	&		&	$B_0=0$	&	$B_0=0.05$	&	$B_0=0.10$	&	$B_0=0.15$	&	$B_0=0.20$	\\\hline
0	&	$\infty$	&	0	&0		&0		&0		&0		\\\hline
4	& 2	& -6.6597	& -6.6944	& -6.7014	& -6.7039	& -6.7051 \\
& 4	& -1.5275	& -1.5406	& -1.5537	& -1.5637	& -1.5709 \\
& 6	& -0.5902	& -0.5940	& -0.5990	& -0.6027	& -0.6055 \\
& 8	& -0.2669	& -0.2684	& -0.2709	& -0.2728	& -0.2743 \\
& 20	& -0.0053	& -0.0054	& -0.0055	& -0.0056	& -0.0057 \\\hline
8	& 2	& -11.6680	& -11.6680	& -11.6650	& -11.6630	& -11.6610 \\
& 4	& -2.7181	& -2.7174	& -2.7351	& -2.7581	& -2.776 \\
& 6	& -1.0665	& -1.0734	& -1.0865	& -1.0967	& -1.1041 \\
& 8	& -0.4960	& -0.5001	& -0.5065	& -0.5113	& -0.5149 \\
& 20	& -0.0105	& -0.0107	& -0.0109	& -0.0112	& -0.0113 \\\hline
16	& 2	& -18.3170	& -18.2970	& -18.2870	& -18.2800	& -18.3260 \\
& 4	& -4.2938	& -4.2845	& -4.3008	& -4.3371	& -4.3665 \\
& 6	& -1.7149	& -1.7187	& -1.7416	& -1.7603	& -1.7740 \\
& 8	& -0.8248	& -0.8322	& -0.8453	& -0.8552	& -0.8623 \\
& 20	& -0.0203	& -0.0207	& -0.0213	& -0.0217	& -0.0220 \\\hline
\end{tabular}
\end{table}
The values of $n^\ast_{\text{min}}$ are strongly influenced by decreasing $K$ (i.e., thickening of EDL) and by increasing $S$. However, the influence of $B_0$ is low to moderate depending on the combination of $K$ and $S$. 
The $n^\ast_{\text{min}}$ values are increasing and approaching to zero as $K$ is increasing from 2 to 20, irrespective of $B_0$ and $S$. Further,  the slip length of 5\% changes $n^\ast_{\text{min}}$ values by (0.52\% and 0.11\%) at $K=2$ and (1.36\% and 1.9\%) at $K=20$ for ($S=4$ and 16). The corresponding changes with slip length of 20\% are noted as (0.68\% and 0.05\%) at $K=2$ and (7.23\% and 8.27\%) at $K=20$. 
\noindent The functional dependence of the minimum value of the excess charge ($n^\ast$, \tab\ref{tab:2}), over the horizontal centreline ($x,0$)  of the microfluidic device, on the dimensionless governing parameters ($K$, $S$ and $B_0$) can be expressed by the following predictive correlation.
\begin{gather}
n^\ast_{\text{min}} = A_1 + A_2 B_0 + A_3 (1/S)+A_4 (B_0/S) + A_5S^2
\label{eq:13}
\\\text{where}\qquad 
A_{\text{i}} = \sum_{{j}=1}^5 M_{\text{ij}} X^{({j}-1)}\quad\text{and}\quad X = (1/K^{1.49})\nonumber
\end{gather}
The correlation coefficients are statistically obtained, for 75 data points, as 
\begin{gather*}
M = \begin{bmatrix}
0.0187	&	2.0248	&	-416.38	&	1951.3	&	-3107.8	\\
0.039	&	-3.2072	&	-101.49	&	1082.4	&	-2150.2	\\
-0.0584	&	-2.8933	&	910.56	&	-4293.1	&	6842.9	\\
-0.1631	&	13.875	&	322.93	&	-4189.7	&	8794.9	\\
0.0003	&	-0.0277	&	-0.2034	&	0.2797	&	-0.1379	
\end{bmatrix} 
\end{gather*}
with $\delta_{\text{min}}=-1.88\%$, $\delta_{\text{max}}=1.20\%$, $\delta_{\text{avg}}=-0.34\%$ and $R^2=99.99\%$ for the range of the conditions explored herein.

%
\subsection{Induced electric field ($E_{\text{x}}$)}
%
\noindent 
In the electroviscous flows (EVF), the electrical field is induced due to the convective transport of the ions in the charged microfluidic device. The electrical field strength ($E_{\text{x}}$) relates (\eqn\ref{eq:A.3}) the total potential ($U$) and EDL potential ($\psi$), and thus, $E_{\text{x}}= - (\partial U/\partial x)$. Further, the conservation of the induced current (\eqn\ref{eq:A.10}) allows to determine the total axial potential gradient ($\nabla U$) due to zero diffusion current ($I_{\text{d}}=0$) at the steady state condition \citep{bharti2009electroviscous}. 
\begin{figure}[h]
	\centering\includegraphics[width=1\linewidth]{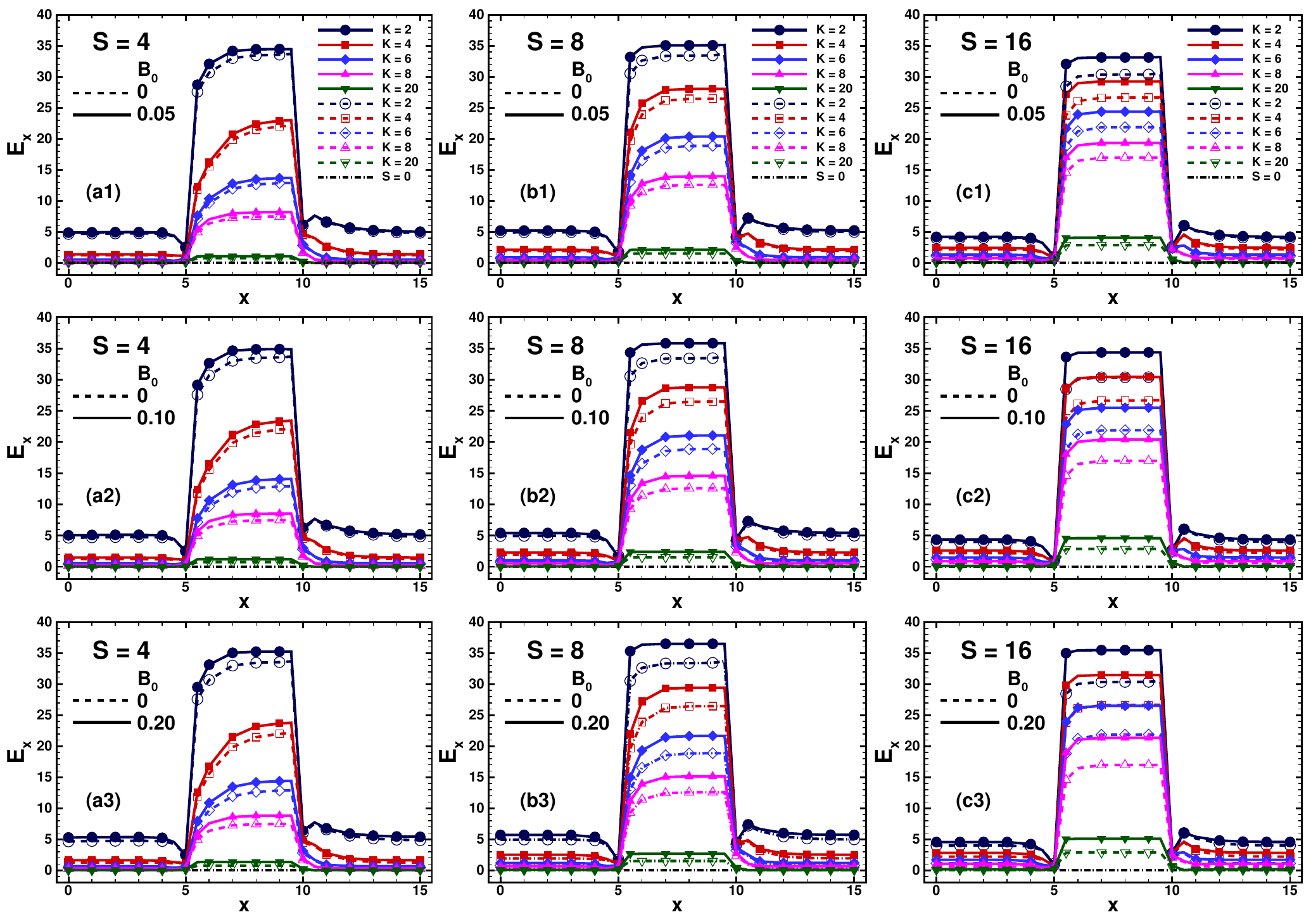}
	\caption{Axial variation of dimensionless induced electric field ($E_{\text{x}}$) in the microfluidic device as a function of dimensionless parameters ($K$, $S$ and $B_0$).}
	\label{fig:11}
\end{figure} 
\fig\ref{fig:11} displays the axial variation of the induced electrical field ($E_{\text{x}}$) as a function of the dimensionless parameters ($K$, $S$ and $B_0$). The dashed ($---$) and solid (\rule[0.5ex]{1cm}{0.5pt}) lines in \fig\ref{fig:11} represent for the no-slip ($B_0=0$) and slip ($B_0>0$) flow conditions. Qualitatively, the electrical field variation over the length has shown similar trends for the ranges of explored conditions.
For instance, the uniform electrical field strength in the inlet section has reduced before the contraction section. In the contraction section, $E_{\text{x}}$ shows sudden shoot up at the entrance, followed by monotonous increases in the first half and negligible increases in the latter half, and then sudden drops at the end.  Further, there is a sudden rise in the starting of the downstream (outlet) section followed by a slow reduction in the first half and attains the constant value in the rest of the outlet section. Furthermore, the field strength in contraction section is excessively higher, due to higher excess charge, in comparison to that in inlet/outlet sections. 
The increase/decrease in $E_{\text{x}}$ in the close vicinity of entrance/exit of the contraction section is primarily attributed to the changes in the field direction due to vertical walls at the end of the inlet and start of the outlet sections. Further, the boundary condition (\eqn\ref{eq:10}) maintain the total electrical potential gradient normal to wall. Altogether, the EDL layer destabilizes due to ions rearrangement and balancing and results in strong fluctuation of $E_{\text{x}}$. The slip ($B_0>0$), however, assists the flow, and thereby the transport of more anions enhances the charge, in turn, the stronger induced electric field is seen in comparison to that under no-slip flow. 
Also, the slip effects enhance with decreasing $K$ and increasing $S$. Irrespective of the slip intensity ($B_0\ge 0$), the induced field strength increases with decreasing $K$ (i.e., thickening of EDL) and with increasing charge density ($S$). 

\begin{table}
\centering\caption{Maximum values of induced electric field strength ($E_{\text{x,max}}$) in the microfluidic device.}\label{tab:emax}
\begin{tabular}{|r|r|r|r|r|r|r|}
\hline
$S$	&	$K$	&	\multicolumn{5}{c|}{$E_{\text{x,max}}$}	\\\cline{3-7}
	&		&	$B_0=0$	&	$B_0=0.05$	&	$B_0=0.10$	&	$B_0=0.15$	&	$B_0=0.20$	\\\hline
0	&	$\infty$	&	0	&0		&0		&0		&0		\\\hline
4	& 2	& 33.6790	& 34.4660	& 34.8770	& 35.1110	& 35.2620 \\
& 4	& 22.1080	& 23.0120	& 23.4110	& 23.6360	& 23.7810 \\
& 6	& 12.9000	& 13.7130	& 14.0770	& 14.2850	& 14.4190 \\
& 8	& 7.4680	& 8.1876	& 8.5130	& 8.6990	& 8.8192 \\
& 20	& 0.7630	& 1.0741	& 1.2155	& 1.2964	& 1.3488 \\\hline
8	& 2	& 33.7070	& 35.1320	& 35.8330	& 36.2460	& 36.5180 \\
& 4	& 26.4830	& 28.0530	& 28.7660	& 29.1710	& 29.4310 \\
& 6	& 18.8840	& 20.3850	& 21.0590	& 21.4420	& 21.6880 \\
& 8	& 12.5870	& 13.9570	& 14.5750	& 14.9260	& 15.1530 \\
& 20	& 1.5062	& 2.1230	& 2.4029	& 2.5628	& 2.6662 \\\hline
16	& 2	& 30.5860	& 33.1800	& 34.3690	& 35.0490	& 35.4930 \\
& 4	& 26.6960	& 29.2600	& 30.4170	& 31.0700	& 31.4880 \\
& 6	& 21.8890	& 24.3850	& 25.4980	& 26.1250	& 26.5270 \\
& 8	& 16.9910	& 19.3450	& 20.3950	& 20.9880	& 21.3670 \\
& 20	& 2.8667	& 4.0587	& 4.5949	& 4.8997	& 5.0963 \\\hline
\end{tabular}
\end{table}

\noindent 
\tab\ref{tab:emax} shows the maximum value of the induced electrical field strength {($E_{\text{x}}$)} as a function of dimensionless parameters ($K$, $S$, $B_0$). While $E_{\text{x}}=0$ for non-electroviscous flows, the minimum value of $E_{\text{x,max}}$ is obtained as 0.7630 at $K=20$ and $S=4$ under no-slip ($B_0=0$) condition. For the given values of $S$ and $B_0$, the electrical field intensity ($E_{\text{x}}$) increases with decreasing value of $K$. For example, the value of $E_{\text{x,max}}$ increases (from 0.7630 to 33.6790), (from 1.5062 to 33.7070) and (from 2.8667 to 30.5860) with decrease in $K$ (from 20 to 2) at $S=4$, 8 and 16, respectively, for no-slip ($B_0=0$) condition. Further, the induced electrical field strength ($E_{\text{x}}$) strengthen with increasing slip ($B_0$). For instance, at $S=4$, $E_{\text{x,max}}$ increased by 2.34\% and 4.7\% with slip ($B_0$) of 5\% and 20\% at $K=2$; the corresponding changes at $K=20$ are 40.77\% and 76.77\%. Similarly, at $S=16$,  $E_{\text{x,max}}$ increased by 8.48\% and 16.04\% with slip ($B_0$) of 5\% and 20\% at $K=2$; the corresponding changes at $K=20$ are 41.58\% and 77.78\%.
\noindent 
The functional dependence of the maximum value of the induced electrical field strength ($E_{\text{x}}$, \tab\ref{tab:emax}) in the microfluidic device, on the dimensionless governing parameters ($K$, $S$ and $B_0$) can be expressed by the following prediction correlation.
\begin{gather}
E_{\text{x,max}} = \sum_{i=1}^{4} A_{\text{i}} (\ln K)^{(i-1)}
\label{eq:14}
\\
\text{where}\qquad  A_{\text{i}} = \sum_{j=1}^{3} A_{\text{ij}} B_0^{(j-1)/2}
\quad \text{and}\quad
A_{\text{ij}} = \sum_{k=1}^{3} M_{\text{ijk}}S^{(k-1)}\nonumber
\end{gather}
The correlation coefficients ($M_{\text{ijk}}$) are statistically obtained, for 75 data points, as
\begin{gather*}
M = \begin{bmatrix} M_1 & M_2 & M_3 & M_4\end{bmatrix}^T
\\
M_1 = \begin{bmatrix}
46.789	&	-4.3668	&	0.181	\\
-3.0307	&	0.5249	&	0.0361	\\
5.2783	&	0.1056	&	-0.0454		
\end{bmatrix}, 
\qquad 
M_2 = \begin{bmatrix}
-13.515	&	9.4724	&	-0.4549	\\
8.6013	&	-0.0148	&	-0.0577	\\
-12.213	&	0.3917	&	0.0402		
\end{bmatrix}, 
\\
\qquad
M_3 = \begin{bmatrix}
-12.716	&	-3.9363	&	0.2298	\\
-5.6406	&	0.4172	&	0.0168	\\
7.3799	&	-0.6017	&	-0.0069	
\end{bmatrix},
\qquad
M_4 = \begin{bmatrix}
4.0109	&	0.4281	&	-0.0328	\\
1.0362	&	-0.142	&	-0.0006	\\
-1.2995	&	0.1484	&	-0.0005		
\end{bmatrix}
\end{gather*}
with $\delta_{\text{min}}=-2.38\%$, $\delta_{\text{max}}=3.22\%$, $\delta_{\text{avg}}=-0.34\%$ and $R^2=99.98\%$ for the range of the conditions explored herein.
%
The preceding discussion have shown the stronger dependence of the total electrical potential ($U$), excess charge ($n^{\ast}$) and induced electric field strength ($E_{\text{x}}$) on the dimensionless parameters ($K$, $S$, $B_0$). As a result, the pressure ($P$) field is also expected to  alter, which is presented and analyzed in the next section. 
%
\subsection{Pressure ($P$)}
\begin{figure}[!h]
	\centering\includegraphics[width=1\linewidth]{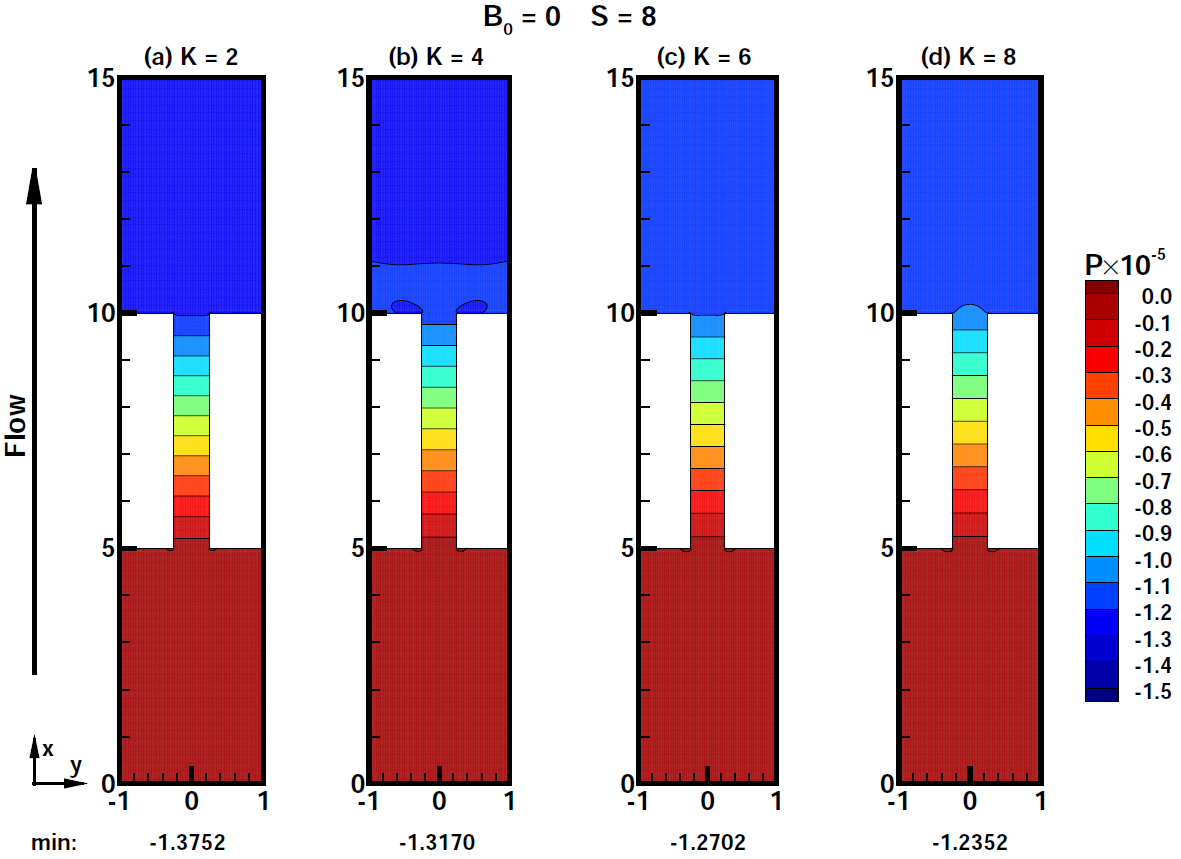}
	\caption{Dimensionless pressure ($P$) distribution as a function of $K$ at $S=8$ and $B_{\text{0}}=0$.}
	\label{fig:8}
\end{figure}

\begin{figure}[h]
	\centering\includegraphics[width=1\linewidth]{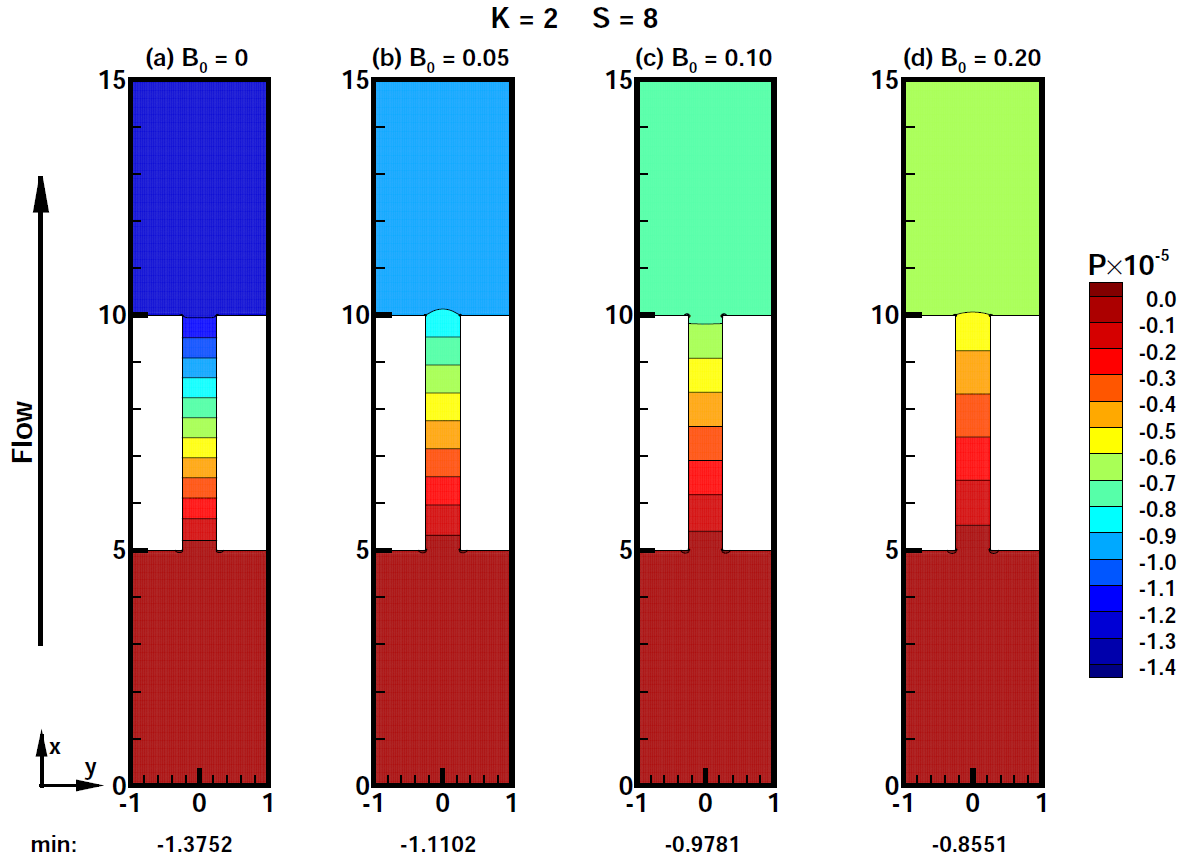}
	\caption{Dimensionless pressure ($P$) distribution as a function of $B_{\text{0}}$ at $S=8$ and $K=2$.}
	\label{fig:9}
\end{figure}

\noindent 
\figs\ref{fig:8} and \ref{fig:9} depict the distribution of the dimensionless pressure ($P$) in the microfluidic device as a function of the  governing parameters ($K$, $S$ and $B_0$).
\fig\ref{fig:8} shows the dependence of the pressure distribution on the inverse Debye length ($2\le K\le 8$) at fixed $S=8$ under the no-slip ($B_{\text{0}}=0$) condition. Qualitatively similar pressure distribution profiles are observed for the other flow conditions, thus not presented here.  The pressure decrease\rev{s}, as expected, over the length of the device. However, the pressure gradient is maximum in the contraction section compared to the upstream and downstream sections of the microfluidic device due to the additional resistance imposed by the excessive charge in the suddenly constricted flow area in the contraction. The reduction in cross-section area leads to an increase in both convective velocity in contraction, compared to upstream and downstream sections, and clustering of the charge occurred in the contraction section due to overlapping of the EDL. So, the convective flow of the excessive ions produced maximum induced electric field (streaming current) and further streaming potential, which imposes additional resistance to the flow that retarded the primary flow in the suddenly constricted flow area.
On the other hand, \fig\ref{fig:9} depicts the influence of the slip length ($0\le B_{\text{0}}\le 0.20$) on the pressure distribution in the device for fixed values of $K=2$ and $S=8$. Qualitatively similar pressure influences are observed for the other flow conditions, thus not presented here. Since the wall slip reduces the wall stress and thereby assists for the flow, the \rev{magnitude of} pressure drop decreases with increasing $B_{\text{0}}$ from $0$ to $0.20$. For instance, the \rev{magnitude of} pressure drop (\rev{$|\Delta P\times 10^{-5}|$}) changes from ${1.3752}$ to ${0.8551}$ with increasing slip ($B_{\text{0}}$) effects from  from 0 to 0.20 at $S=8$ and $K=2$.

\begin{figure}[h]
	\centering\includegraphics[width=1\linewidth]{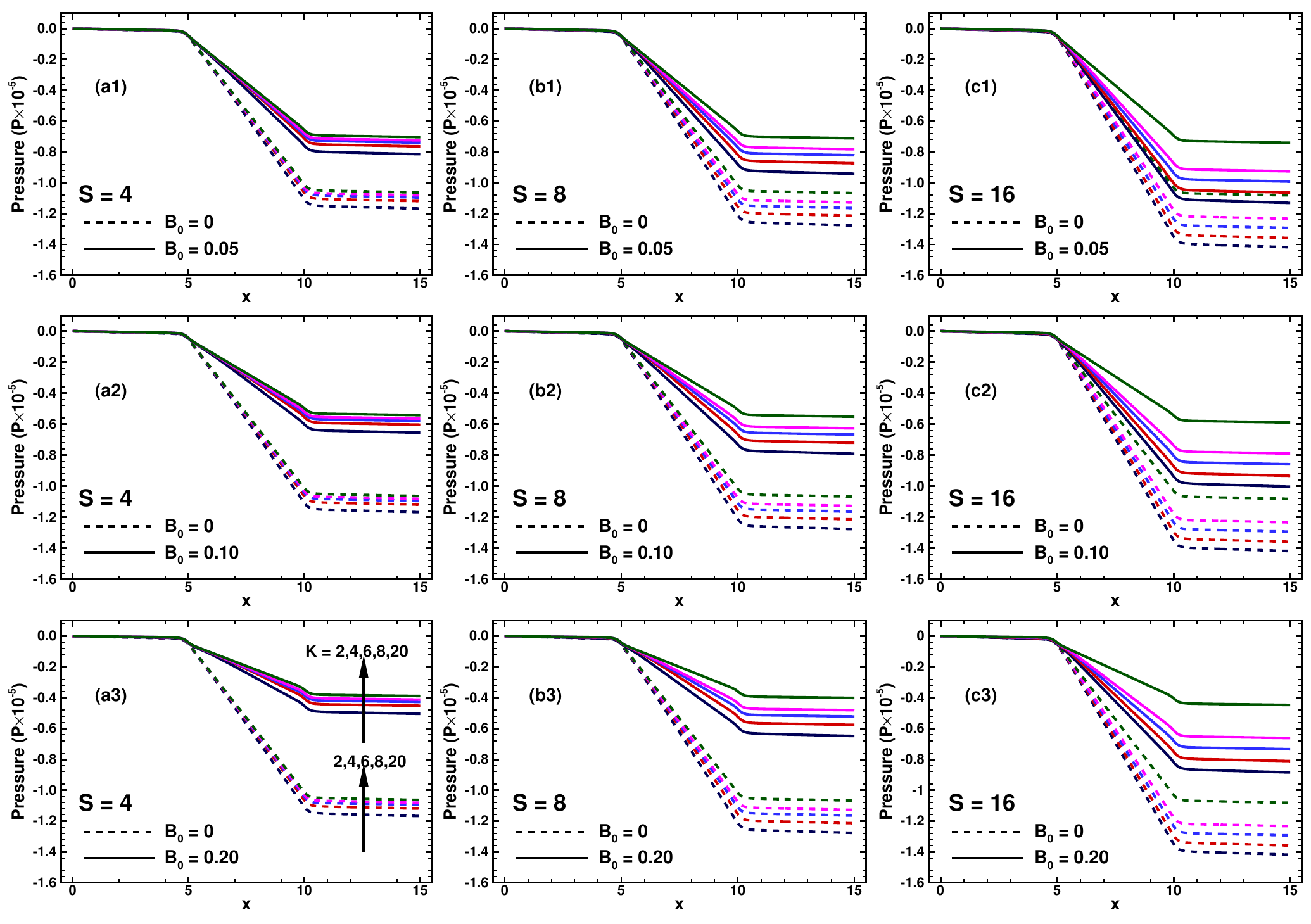}
	\caption{Axial variation of dimensionless pressure ($P$)  along the horizontal centreline ($x,0$) of the microfluidic device as a function of dimensionless parameters ($K$, $S$ and $B_0$).}
	\label{fig:10}
\end{figure}

\noindent 
Further combined influences of dimensionless parameters ($K$, $S$, and $B_0$) are gained in \fig\ref{fig:10} through axial variation of the pressure along the horizontal centreline ($x,0$) of the device for $2\le K\le 20$, $4\le S\le 16$ and $0\le B_0\le 0.20$. The dashed ($---$) and solid (\rule[0.5ex]{1cm}{0.5pt}) lines in \fig\ref{fig:10} represent for the no-slip ($B_0=0$) and slip ($B_0>0$) flow conditions. 
The pressure ($P$) profiles are qualitatively similar to the electrical potential ($U$, \fig\ref{fig:6}) profiles, i.e., the pressure decreases axially along the length of the device for all conditions. The maximum drop ($\Delta P$) is obtained in the contraction section, irrespective of the flow conditions.  While the pressure variation is qualitatively consistent for both slip and no-slip flow, quantitative influences of $B_0$ are notable in \fig\ref{fig:10}. The pressure drop ($\Delta P$) increases with increasing slip length ($B_0$). The pressure drop \rev{decreases} by about $63\%$ \rev{at fixed $K$ and $S$ when $B_0$ is increased from $0$ to $0.20$,} over the broad ranges of conditions. \rev{The maximum change obtains at $K=20$ and $S=4$ when electroviscous effects are weakest. It is because at smaller $S$ and higher $K$, additional resistance applied by the streaming potential is lesser (as shown in \fig\ref{fig:6} and \tab\ref{tab:1}), therefore, reduction in the pressure drop is maximum with slip length when electroviscous effects are weakest.} 

\noindent
A detailed analysis of pressure drop in charge-dependent slip flow of electrolyte liquids is subsequently presented. 
%
\noindent 
\tab\ref{tab:4} display the variation of pressure drop ($|\Delta P| \times 10^{-5}$) over the length of device with the dimensionless flow parameters ($K$, $S$ and $B_0$). \tab\ref{tab:4}  also includes the data for non-electroviscous ($S=0$ or $K=\infty$) flows.
As discussed earlier, the pressure drop decreases with the increasing value of $K$, irrespective of $S$ and $B_0$. 
It is also noted that the increasing slip intensity ($B_0$) reduces the pressure drop ($|\Delta P|$), and increasing surface charge density ($S$) enhances the pressure drop \rev{($|\Delta P|$)}. Further, the minimum pressure drop ($|\Delta P|_{\text{min}}$) is obtained at largest $K=20$ for all $S$ and $B_0$. 
\begin{table}[h]
\centering\caption{Slip effects ($0\le B_0\le 0.20$) on the pressure drop ($10^{-5}|\Delta P|$) over the length of the microfluidic device in electroviscous ($2\le K\le 20$ and $4\le S\le 16$) and non-electroviscous ($S=0$ or $K=\infty$) flows.}\label{tab:4}
\begin{tabular}{|r|r|r|r|r|r|r|}
\hline
$S$	&	$K$	&	\multicolumn{5}{c|}{$10^{-5}|\Delta P|$}	\\\cline{3-7}
	&		&	$B_0=0$	&	$B_0=0.05$	&	$B_0=0.10$	&	$B_0=0.15$	&	$B_0=0.20$	
	\\\hline
0	&	$\infty$	&	1.0616	&	0.7013	&	0.5384	&	0.4453	&	0.3848	
	\\\hline
4	& 2	& 1.1673	& 0.8141	& 0.6545	& 0.5634	& 0.5043 \\
& 4	& 1.1189	& 0.7635	& 0.6030	& 0.5112	& 0.4517 \\
& 6	& 1.0952	& 0.7394	& 0.5787	& 0.4867	& 0.4271 \\
& 8	& 1.0811	& 0.7247	& 0.5638	& 0.4718	& 0.4121 \\
& 20	& 1.0629	& 0.7039	& 0.5418	& 0.4491	& 0.3889 \\\hline
8	& 2	& 1.2767	& 0.9412	& 0.7902	& 0.7041	& 0.6483 \\
& 4	& 1.2136	& 0.8733	& 0.7200	& 0.6324	& 0.5757 \\
& 6	& 1.1640	& 0.8213	& 0.6669	& 0.5787	& 0.5215 \\
& 8	& 1.1274	& 0.7824	& 0.6270	& 0.5383	& 0.4807 \\
& 20	& 1.0669	& 0.7116	& 0.5517	& 0.4603	& 0.4010 \\\hline
16	& 2	& 1.4180	& 1.1304	& 1.0030	& 0.9309	& 0.8845 \\
& 4	& 1.3579	& 1.0635	& 0.9327	& 0.8585	& 0.8106 \\
& 6	& 1.2929	& 0.9922	& 0.8585	& 0.7826	& 0.7337 \\
& 8	& 1.2329	& 0.9256	& 0.7890	& 0.7115	& 0.6615 \\
& 20	& 1.0815	& 0.7407	& 0.5888	& 0.5025	& 0.4467 \\\hline
\end{tabular}
\end{table}
The lowest values of pressure drop ($|\Delta P|$) are obtained quite close to that for non-electroviscous ($S=0$ or $K=\infty$) flow, under otherwise identical conditions, at smaller $S$ and $B_0$. For instance, under no-slip ($B_0=0$) condition, $|\Delta P|$ is higher by 0.12\%, 0.5\% and 1.87\%  for $S=4$, 8 and 16, respectively, at $K=20$ compared to $|\Delta P|=1.0616$ for $K=\infty$. 
On increasing $K$ from 2 to 20, $|\Delta P|$ reduced by 8.94\%, 16.43\% and 23.73\% at $S=4$, 8 and 16, respectively, at $B_0=0$.
Further, $|\Delta P|$ reduced from 1.0616 (at $B_0=S=0$) by 33.94\%, 49.28\%, 58.05\% and 63.75\% with increasing slip ($B_0$) intensity as 5\%, 10\%, 15\% and 20\%, respectively. 

\noindent 
The influence of slip ($B_0$) intensity on the pressure drop are stronger in weakly electroviscous or non-electroviscous flows in comparison to that in strongly electroviscous flows. For example, the pressure drop \rev{($|\Delta P|$)} reduces by 56.8\%, 49.22\% and 37.62\% with increases in $B_0$ from 0 to 20\% for $S=4$, 8, and 16, respectively, at $K=2$; the corresponding reduction in $|\Delta P|$ at $K=20$ is noted as 63.41\%, 62.41\% and 58.7\%.  
On the other hand, the pressure drop \rev{($|\Delta P|$)} increased by (9.82\%, 19.66\% and 31.11\%) and (29.66\%, 61.65\% and 98.01\%) at ($S =$ 4, 8, and 16) for $B_0=0$ and 20\% with decreasing $K$ from 20 to 2.
Overall, the pressure drop ($|\Delta P|$) has shown a complex interplay between dimensionless parameters ($K$, $S$, and $B_0$). An increasing slip intensity ($B_0$) assists the flow, weakens the resistance, and reduces the pressure drop ($|\Delta P|$). However, the flow resistance enhances with the thickening of EDL (i.e., decreasing $K$) and an increasing surface charge density ($S$), resulting in an increased pressure drop ($|\Delta P|$). 

\noindent
The preceding discussion has shown that the flow characteristics such as electrical potential, excess charge, electrical field, and pressure in the microfluidic device intricately depend on the flow governing parameters.
%
\subsection{Electroviscous correction factor ($Y$)}
\noindent
In the electrokinetic flows, additional flow resistance is imposed by the streaming potential on the charged liquid in EDL. The streaming potential generates from the electric field induced by the convective migration of the excess charge near the charged walls. Consequently, this additional flow resistance manifests the pressure drop ($\Delta P$) along with the device, which is higher than the pressure drop without electrical forces  ($\Delta P_0$) for a fixed volumetric flow rate.
The relative enhancement of the pressure due to the induced electrical field is generally measured in terms of an apparent or effective viscosity ($\mu_{\text{eff}}$) and referred to as the electroviscous effects \citep{davidson2007electroviscous,davidson2008electroviscous, bharti2008steady,bharti2009electroviscous}. The effective viscosity ($\mu_{\text{eff}}$) is the viscosity of the fluid, in the absence of the electrical force, needed to obtain the pressure drop ($\Delta P$).

\noindent 
For the low Reynolds number ($Re$) laminar steady microfluidic flow, the nonlinear advection term is negligibly small in the momentum equation (\eqn\ref{eq:A.6}), that is, the left side of \eqn(\ref{eq:A.6}) becomes negligible. The relative enhancement in the pressure drop ($\Delta P/\Delta P_0$) is, thus, attributed to the correspondingly relatively higher viscosity ($\mu_{\text{eff}}/\mu$) of the fluid, under otherwise identical conditions. These relative quantities define the electroviscous effects as the \textit{electroviscous correction factor} ($Y$) expressed as follows for  a given slip length ($B_0$).
\begin{gather}
	Y=\frac{\mu_{\text{eff}}}{\mu}=\frac{\Delta P}{\Delta P_{\text{0}}}
	\label{eq:15}
\end{gather}
where, $\Delta P$ and $\Delta P_0$ are the pressure drop with electroviscous ($S>0$) effects and the pressure drop without electroviscous ($S=0$ or $K=\infty$) effects, respectively. The subscript `0' represents the quantity without electroviscous effects, i.e., in the absence of an electric field.
Further, the variables $\mu$ and $\mu_{\text{eff}}$ denote the viscosities of the liquid yielding the pressure drop ($\Delta P_0$) and effective pressure drop ($\Delta P_{\text{eff}}=\Delta P$), both in the absence of the electric field.
\begin{figure}[!h]
	\centering\includegraphics[width=0.75\linewidth]{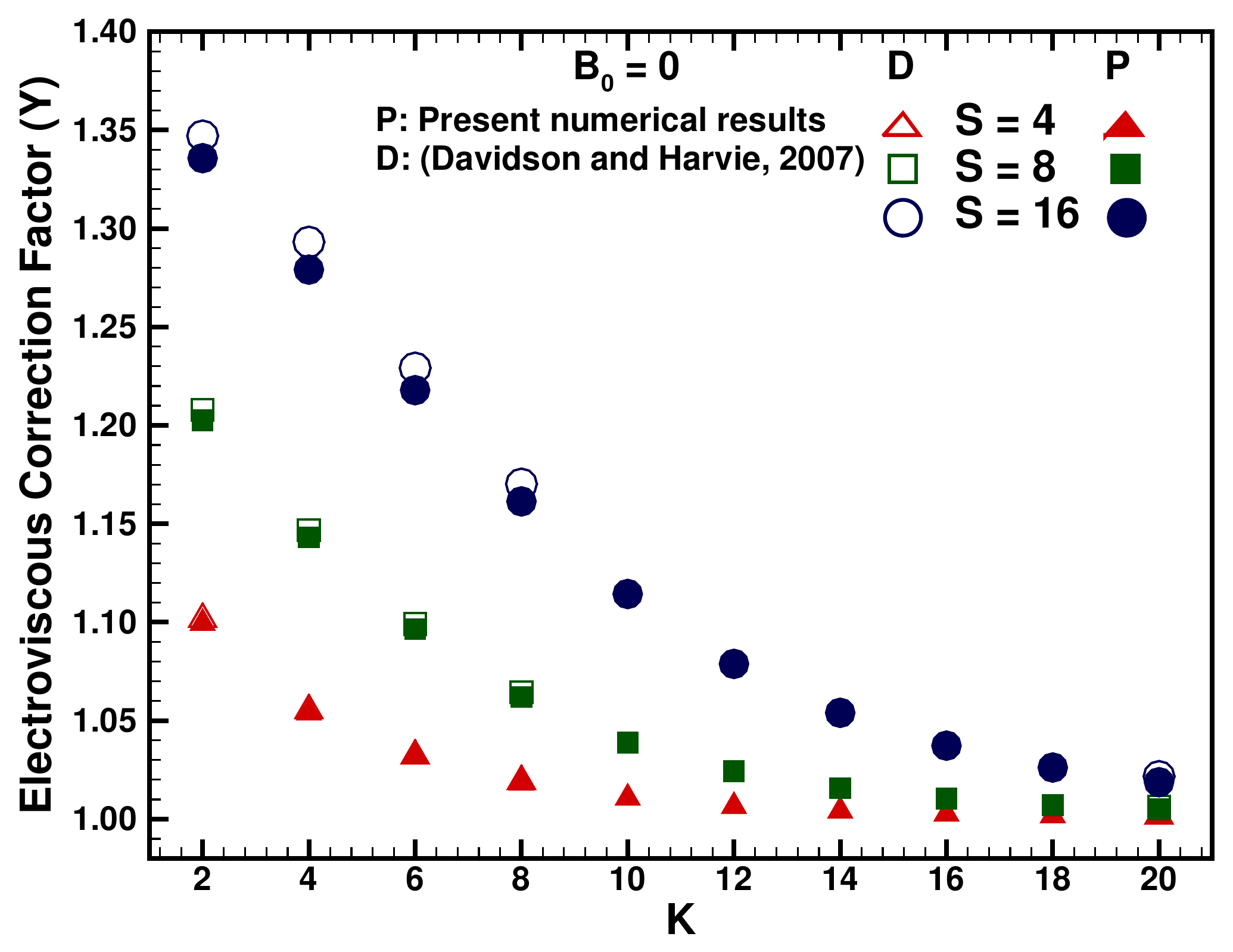}
	\caption{Comparison of the present and literature values of electroviscous correction factor ($Y$) under no-slip ($B_{\text{0}}=0$) condition.}
	\label{fig:12}
%
	\centering\includegraphics[width=0.75\linewidth]{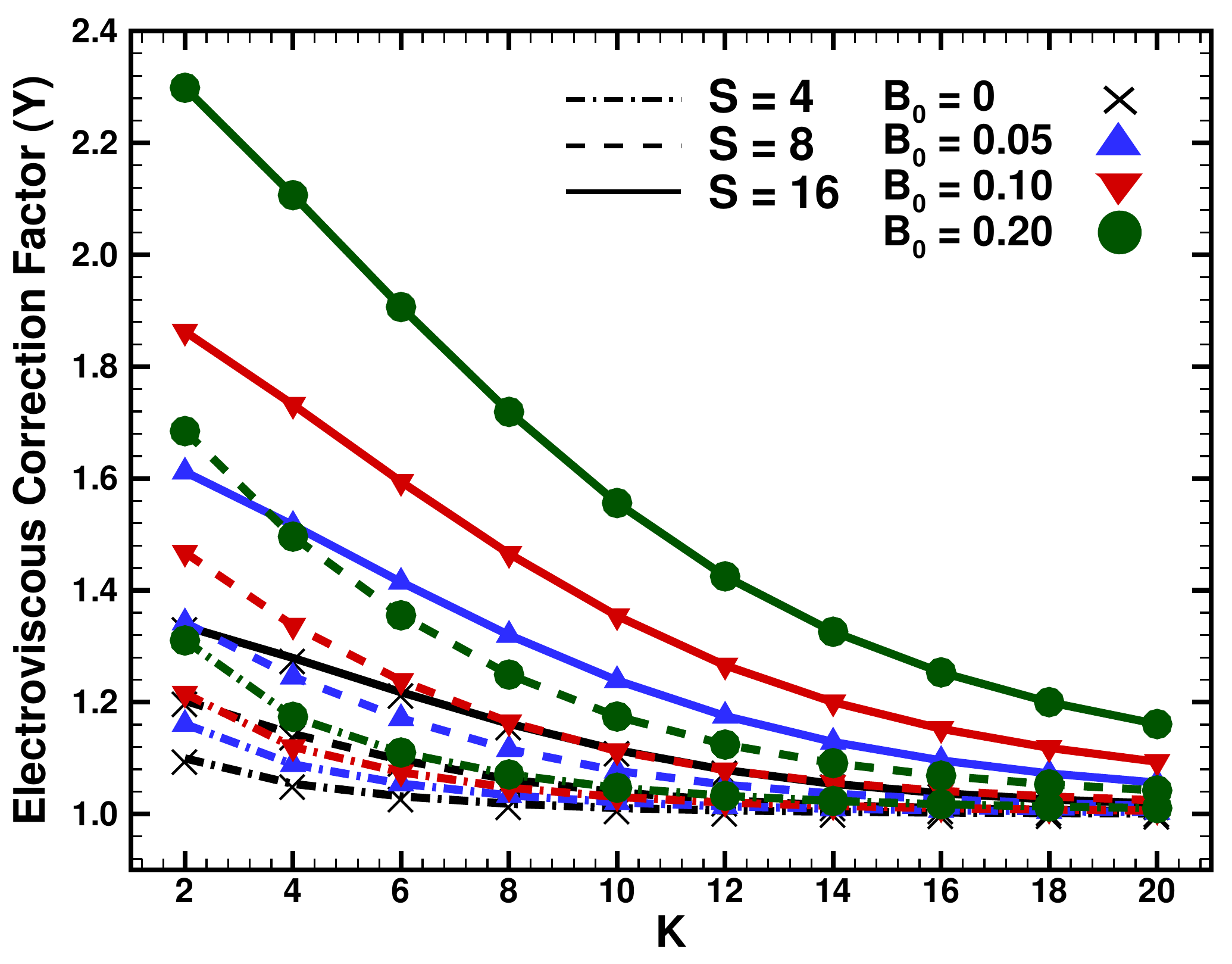}
	\caption{Electroviscous correction factor ($Y$) as a function of flow governing parameters ($K$, $S$ and $B_{\text{0}}$).}
	\label{fig:13}
\end{figure} 

\noindent 
\fig\ref{fig:12} compares the present and literature \citep{davidson2007electroviscous} values of electroviscous correction factor {($Y$)} in the limits of no-slip ($B_0=0$) electroviscous ($2\le K\le 20$; $4\le S\le 16$) flow conditions. Both results have shown excellent agreement, even when obtained using different numerical approaches.

\noindent
Subsequently, \fig\ref{fig:13} displays the variation of the electroviscous correction factor with the flow governing parameters ($K$, $S$ and $B_0$) for charge-dependent slip flow of electrolyte liquid flow through a charged microfluidic device. Qualitative variations of $Y$ with $K$ and $S$ are similar to both slip ($B_0>0$) and no-slip ($B_0=0$) conditions (\fig\ref{fig:13}). For instance, the electroviscous correction factor is above one ($Y>1$) for the ranges of conditions explored. The electroviscous correction factor has shown proportional dependence on both surface charge density ($S$) and slip intensity ($B_0$); however, inversely proportional dependence on Debye parameter ($K$). This trend of $Y$ suggests strengthening of the electroviscous effects with increasing $B_0$. Due to increased axial potential gradient increases, with increasing $B_0$, (as shown in \fig\ref{fig:6} and \tab\ref{tab:1}) an extra body force exerted on the EDL is greater. Further, $Y$ increases with increasing $S$ \rev{due to increment in the pressure drop with increasing $S$ (as shown in \fig\ref{fig:10} and \tab\ref{tab:4}).} \rev{The electroviscous correction factor increases with} decreasing $K$ or EDL thickening. \rev{It is} because the streaming current \rev{(and hence $Y$) is maximum for small enough $K$. This maximum value occurs once EDL strongly overlapped, thus no further increase in the rate of charge advection (i.e., the streaming current, by definition) is possible}.
For the ranges of conditions presented here, the electroviscous correction factor  increases maximally by  33.57\% at smallest $K=2$ and largest $S=16$ in no-slip ($B_0=0$) flow.  The corresponding maximum increase in $Y$ in the slip ($B_0>0$) flow is recorded as 129.87\% at smallest $K=2$, largest $S=16$  and largest $B_0=0.20$. The influence of slip ($B_0>0$), in comparison to no-slip ($B_0=0$), on $Y$ is also noted as high as 72.1\%. 

\noindent 
In general, the electroviscous correction factor {($Y$)} has shown the complex dependence on the flow governing parameters. 
The functional dependence of the electroviscous correction factor ($Y$) on the dimensionless flow governing parameters ($K$, $S$, $B_0$) is expressed as follows. 
\begin{gather}
Y = A_{1} + A_{2}K + A_{3}S + A_{4}K^2+ A_{5}KS + A_{6} K^3S^2+ A_{7} K^3
\label{eq:16}
\\
\text{where}\qquad 
A_{\text{i}} = \sum_{{j}=1}^3 M_{\text{ij}} B_0^{({j}-1)}\qquad \text{for}\qquad 1\le i \le 7 \nonumber
\end{gather}
where, the correlation coefficients ($M_{\text{ij}}$) are statistically obtained as 
\begin{gather*}
M = \begin{bmatrix}
1.060499	&	-0.0226	&	0.024	&	0.00172	&	-0.00157	&	\num{5.120E-08}	&	-\num{3.27E-05}	\\
0.0108	&	-0.21154	&	0.440486	&	0.017797	&	-0.025	&	\num{7.380E-07}	&	-\num{3.38E-04}	\\
-0.0956	&	0.165714	&	-0.35143	&	-0.01229	&	0.0209	&	-\num{6.170E-07}	&	\num{2.14E-04}	
\end{bmatrix}^{T} 
\end{gather*}
with $\delta_{\text{min}}=-2.09\%$, $\delta_{\text{max}}=2.04\%$, $\delta_{\text{avg}}=-0.12\%$ and $R^2=99.77\%$ for the range of the conditions explored herein.
%
%
\subsection{Pseudo-analytical model}
\noindent
Earlier studies have proposed simple analytical models to predict the pressure drop ($\Delta P$) in no-slip flow through contraction-expansion microchannels of \rev{slit} \citep{davidson2007electroviscous} and circular \citep{bharti2008steady} cross-sections. A similar approach has been used in the present study to propose a simple predictive model to obtain the pressure drop ($\Delta P$) and, hence the electroviscous correction factor ($Y$) for the surface charge-dependent slip flow of the symmetric electrolytes through the \rev{slit} contraction-expansion microfluidic device. The proposed analytical/mathematical model to obtain the total pressure drop ($\Delta P$) in the \rev{slit} contraction-expansion microfluidic device is expressed by \eqn(\ref{eq:17}). 
\begin{gather}
\Delta P_{\text{m}}=(\Delta P_{\text{u}}+\Delta P_{\text{c}}+\Delta P_{\text{d}})+\Delta P_{\text{e}}
\label{eq:17}
\end{gather}
where $\Delta P_{\text{u}}$, $\Delta P_{\text{d}}$ and $\Delta P_{\text{c}}$ denote for the pressure drop in upstream, downstream and contraction sections. 
Notably, these sections individually depict the \rev{slit} microchannels of uniform cross-section. 
The pressure drop ($\Delta P$) in the steady laminar incompressible Newtonian fully-developed Poiseuille flow, in the absence of both slip and electrical field, through the uniform \rev{slit} channel of length ($\Delta L$) can be determined by the standard \textit{Hagen–Poiseuille equation} as follows. 
\begin{gather*}
{\Delta P_{00}}= \left(\frac{3}{Re}\right){\Delta L}
\end{gather*}
\noindent
Further, the excess pressure drop ($\Delta P_{\text{e}}$, \eqn\ref{eq:17}) due to sudden contraction and expansion is approximated by the pressure drop through thin orifices \rev{($d_\text{c}<<1$)} in absence of slip and electrical field \citep{Sisavath2002,davidson2007electroviscous,bharti2008steady,Pimenta2020} and expressed as follows.
\begin{gather}
\Delta P_{00,\text{e}}=\frac{16}{\pi d_{\text{c}}^2Re}
\label{eq:18}
\end{gather}
%
\rev{\eqn(\ref{eq:18}) is applicable for $d_\text{c}<<1$.}
\noindent
In the absence of both slip ($B_0=0$) and electrical field ($S=0$ and $K=\infty$), a generalized mathematical model for the pressure drop in the flow through the contraction-expansion microfluidic device is, thus, expressed as follow.
\begin{gather}
\Delta P_{00,\text{m}}=(\Delta P_{00,\text{u}}+\Delta P_{00,\text{c}}+\Delta P_{00,\text{d}})+\Delta P_{00,\text{e}}
\label{eq:19}
\end{gather}
where,
\begin{gather}
{\Delta P}_{00,\text{u}}= \left(\frac{3}{Re}\right){L_{\text{u}}},\qquad
{\Delta P}_{00,\text{c}}= \left(\frac{3}{d_{\text{c}}^3Re}\right)L_{\text{c}},
\qquad\text{and}\qquad
{\Delta P}_{00,\text{d}}= \left(\frac{3}{Re}\right){L_{\text{d}}}\nonumber
\end{gather}
where the length variables ($L_{\text{u}}$, $L_{\text{d}}$ and $L_{\text{c}}$) are scaled with $W$, and the $Re$ is defined in \eqn(\ref{eq:5}).
Note the typographical inadvertent omission of a factor $(1/{d_{\text{c}}^{3}})$ in the second term accounting for ${\Delta P}_{\text{c}}$ in \eqn (23) of  \citet{davidson2007electroviscous}.

\noindent
Subsequently, \eqn(\ref{eq:19}) is modified to account for slip effects on the pressure drop. In the absence of electrical field ($S=0$ and $K=\infty$), a generalized mathematical model for the pressure drop in the slip ($B_0>0$) flow through the contraction-expansion microfluidic device is, thus, expressed as follow.
\begin{gather}
\Delta P_{0,\text{m}}= \Gamma_0 \Delta P_{00,\text{m}}
=
\left(\frac{3\Gamma_0}{Re}\right)\left(L_{\text{u}} +  \frac{L_{\text{c}}}{d_{\text{c}}^3} + L_{\text{d}} + \frac{16}{3\pi d_{\text{c}}^2} \right)
\label{eq:20}
\end{gather}

\noindent 
The correction coefficient ($\Gamma_0$, \eqn\ref{eq:20}) accounts for influence of slip length ($B_0>0$) on the pressure drop ($\Delta P_{00,\text{m}}$)  as follows. 
\begin{gather}
\Gamma_0 = C_1 + C_2 B_0^{0.5} + C_3 B_0^3
\label{eq:21}
\end{gather}
The correlation coefficients ($C_{\text{i}}$, \eqn\ref{eq:21}) are statistically obtained as 
$C_1 = 0.993403$, $C_2 = -1.55851$, and $C_3 = 7.691732$
with $\delta_{\text{min}}=-1.15\%$, $\delta_{\text{max}}=1.31\%$, $\delta_{\text{avg}}=0.03\%$ and $R^2=99.96\%$ for the range of the conditions explored herein.

\noindent
The mathematical model (\eqn\ref{eq:19}) is further modified to account both slip and electroviscous effects on the pressure drop. 
In the presence of both slip ($B_0>0$) and electrical field ($S>0$ and $0 < K < \infty$), a generalized mathematical model for the pressure drop in the flow of electrolyte liquids through the charged contraction-expansion microfluidic device is, thus, expressed as follow.
\begin{gather}
\Delta P_{\text{m}}= \Gamma \Delta P_{00,\text{m}}
=
\left(\frac{3\Gamma}{Re}\right)\left(L_{\text{u}} +  \frac{L_{\text{c}}}{d_{\text{c}}^3} + L_{\text{d}} + \frac{16}{3\pi d_{\text{c}}^2} \right)
\label{eq:22}
\end{gather}
The correction coefficient ($\Gamma$, \eqn\ref{eq:22}) accounts for influence of both slip ($B_0>0$) and electroviscous ($S>0$) effects on the pressure drop ($\Delta P_{00,\text{m}}$) as follows.
\begin{gather}
\Gamma =   A_1 + A_2 K + A_3 S + 10^{-4}A_4 K^2 + 10^{-4}A_5S^2 + A_6K^{0.5}S 
\label{eq:23}
\\\text{where}\qquad 
A_{\text{i}} = \sum_{j=1}^{4} M_{\text{ij}}B_0^{(j-1)}
\qquad\text{for}\qquad 1\le i \le 6 \nonumber
\end{gather}
The correlation coefficients ($M_{\text{ij}}$, \eqn\ref{eq:23}) are statistically obtained as 
\begin{gather*}
M = \begin{bmatrix}
1.0166	&	-9.1138	&	50.603	&	-106.91	\\
-0.0142	&	0.0435	&	-0.2489	&	0.5328	\\
0.0344	&	0.1151	&	-0.6178	&	1.2869	\\
5.7461	&	-12.238	&	72.971	&	-159.07	\\
-2.112	&	33.12	&	-162.1	&	323.62	\\
-0.0063	&	-0.0318	&	0.1692	&	-0.3516	
\end{bmatrix} 
\end{gather*}
with $\delta_{\text{min}}=-3.94\%$, $\delta_{\text{max}}=2.37\%$, $\delta_{\text{avg}}=-0.80\%$ and $R^2=99.86\%$ for the range of the conditions explored herein.
\begin{figure}[htbp]
	\centering\includegraphics[width=0.75\linewidth]{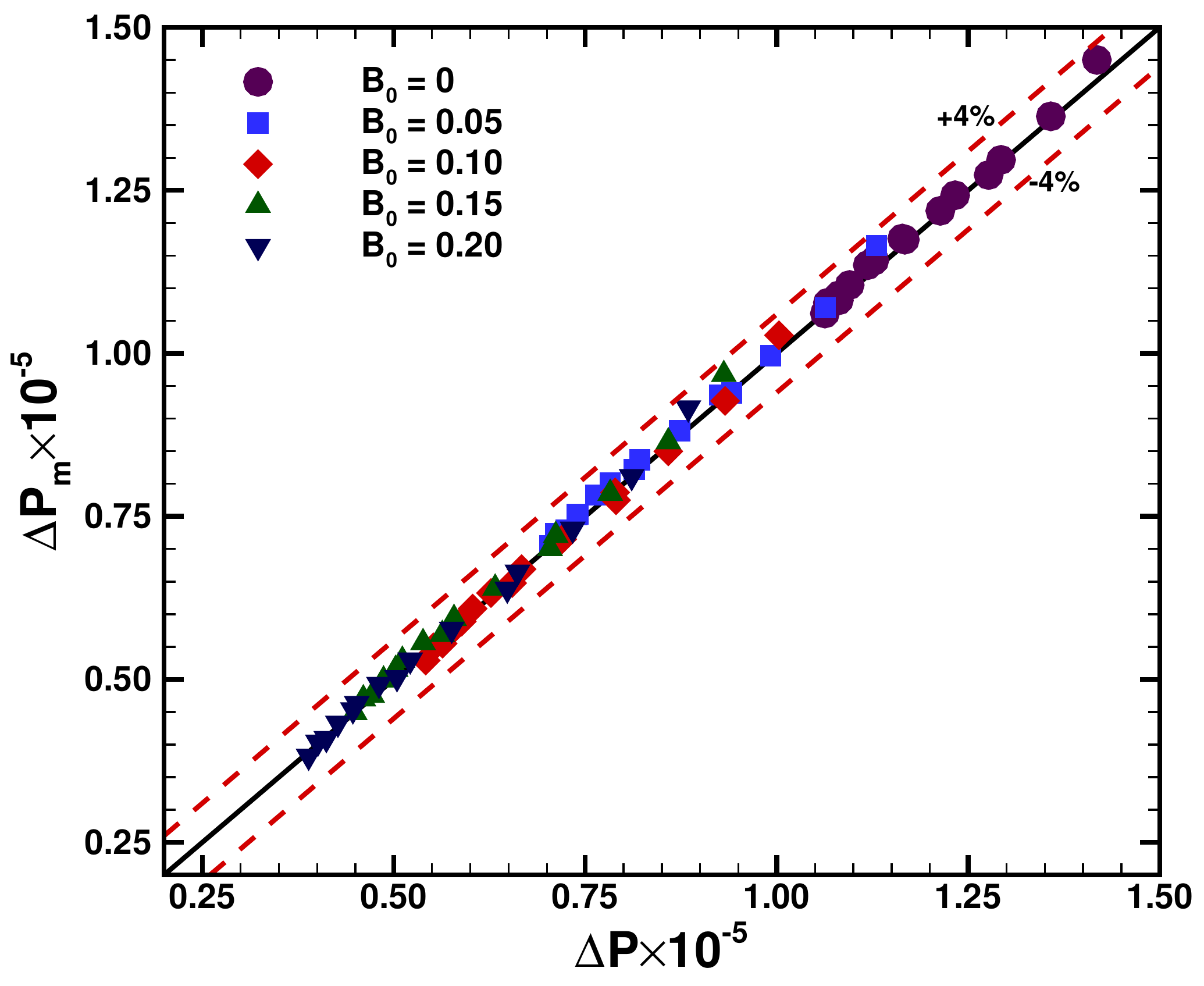}
	\caption{Parity chart for pressure drop values obtained numerically ($\Delta P$) and mathematically ($\Delta P_{\text{m}}$, \eqn\ref{eq:22}) for the considered parameters ($K$, $S$ and $B_0$).}
	\label{fig:14}
\end{figure} 
\begin{figure}[h]
	\centering\includegraphics[width=0.75\linewidth]{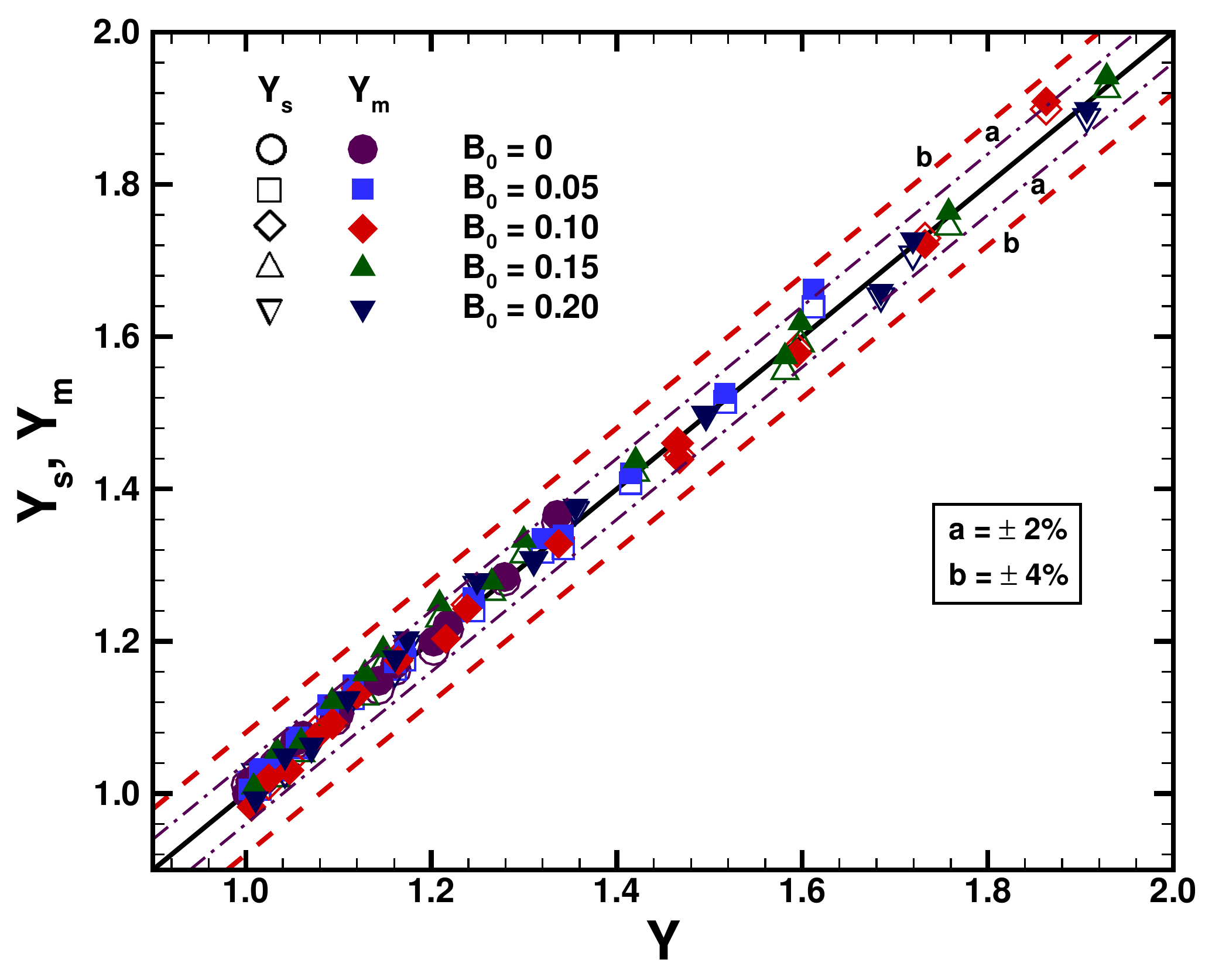}
	\caption{Parity chart for the electroviscous correction factor values obtained numerically ($Y$) and mathematically ($Y_{\text{m}}$, \eqn\ref{eq:24}; $Y_{\text{s}}$, \eqn\ref{eq:16}) for the considered parameters ($K$, $S$ and $B_0$).}
	\label{fig:15}
\end{figure} 

\noindent
The above presented simpler analytical model (\eqns\ref{eq:20} and \ref{eq:22}) for the low Reynolds number flow through the contraction-expansion microfluidic device is further extended to predict the electroviscous correction factor as follows.
\begin{gather}
Y_{\text{m}}=\frac{\Delta P_{\text{m}}}{\Delta P_{0,\text{m}}}
\label{eq:24}
\end{gather}
\noindent
\figs\ref{fig:14} and \ref{fig:15} present the parity charts for the pressure drop ($\Delta P$ vs $\Delta P_{\text{m}}$) and the electroviscous correction factor ($Y$ vs $Y_{\text{m}}$) obtained using the present numerical approach and from the simple predictive mathematical model (\eqn\ref{eq:22} or \ref{eq:24}) for the ranges of conditions ($K$, $S$ and $B_0$) considered in this work.
The simpler model estimates both the pressure drop and electroviscous correction factor within $\pm 2-4\%$ of the numerical values. The difference between numerical simulation and predicted results reduce with decreasing surface charge density and thinning of EDL. Such a simple approach \citep{davidson2007electroviscous,bharti2008steady,bharti2009electroviscous} for the prediction of pressure drop, and thereby, electroviscous correction factor enables the use of present results in the design and engineering of relevant microfluidic devices.
%
\section{Concluding remarks}
\noindent
This study has explored the slip effects in the steady laminar flow of symmetric (1:1) electrolyte through the uniformly charged \rev{slit} contraction-expansion (4:1:4) microfluidic device. The charge-dependent slip is considered at the device walls. The mathematical model equations, including the Poisson’s equation, Navier-Stokes equation with electrical body force, Nernst-Planck equation, and current continuity equation, are solved numerically 
using finite element method (FEM).
Numerically results for total electrical potential, charge, pressure, induced electric field strength, pressure drop and electroviscous correction factor are discussed for the broader ranges of conditions ($2\le K \le 20$, $4\le S\le 16$, $0\le B_{0}\le 0.20$, $\mathit{Sc}=1000$, and $Re=0.01$).
The effect of non-uniformity geometry on the flow fields has been analyzed and found that the sudden contraction/expansion in the geometry tremendously increases the excess charge, induced electric field strength, potential drop and pressure drop in the microfluidic device, irrespective of the governing parameters.
The flow fields have shown complex dependence on the flow governing parameters. \rev{Results show that, over the range of the conditions, the total electrical potential drop ($|\Delta U|$) maximally increases by 78.68\% (at $K=20$, $S=16$, $B_0=0.20$) and  the pressure drop ($|\Delta P|$) maximally decreases by 63.42\% (at $K=20$, $S=4$, $B_0=0.20$) with respect to the no-slip condition.}
The electroviscous correction factor ($Y$) increases maximally by 33.58\% under the no-slip ($B_{\text{0}}=0$) condition. 
Further, the electroviscous correction factor \rev{($Y$) increases maximally by 19.19\% (at $K=2$, $S=4$, $B_0=0.20$)  and 72.10\% (at $K=2$, $S=16$, $B_0=0.20$), respectively,  with respect to the no-slip values, over the range of the conditions.} Thus, the surface charge-dependent slip enhances the electroviscous effect in the microfluidic device than the no-slip flow.
Further, a simpler model is introduced to estimate the pressure drop (and hence electroviscous correction factor) in the microfluidic device by adding the pressure drop for all sections (upstream, contraction, and downstream) separately and excess pressure drop (due to converging and diverging flow areas of the device). The simpler model estimates the pressure drop of $\pm 2-4\%$ compared to the numerical results. The difference between the numerical simulation results and predicted results of the simpler model is becoming negligible when the surface charge density and EDL thickness decrease. A simpler mathematical model enables the use of present results in designing and engineering relevant microfluidic devices. \rev{Further, both charge-dependent slip and non-uniform geometry effects increase the electroviscous impact, i.e., retards the primary pressure-driven flow of liquid and increases the residence time for a fixed length of microchannel. The outcome of the present work, thus, can be utilized to intensify the complex microfluidic electrokinetic transport processes, including mixing, diffusion, heat and mass transfer, reaction.}
%
%
\section*{Declaration of Competing Interest}
\noindent 
The authors declare that they have no known competing financial interests or personal relationships that could have appeared to influence the work reported in this article.
%
\section*{Acknowledgements}
\noindent 
R.P. Bharti would like to acknowledge Science and Engineering Research Board (SERB), Department of Science and Technology (DST), Government of India (GoI) for the providence of the MATRICS grant (File no. MTR/2019/001598).
%
%
%
\begin{spacing}{1.5}
\input{Nomenclature.tex}

\printnomenclature
\end{spacing}
%
%
%
%
%
%
%
\bibliographystyle{elsarticle/elsarticle-harv}\biboptions{authoryear}
\bibliography{references}
%
%
%
%
%
%
%
%
%
%
\end{document}

%% file: nomenclature.tex
\fontsize{10}{10pt}\selectfont
 \nomenclature[g0]{\textit{Greek letters}}{}
 \nomenclature[d0]{\textit{Dimensionless groups}}{}
 \nomenclature[s0]{\textit{Subscripts and Superscripts}}{}
 \nomenclature[z0]{\textit{Abbreviations}}{}
%
\nomenclature[zcfd]{CFD}{computational fluid dynamics}
\nomenclature[zedl]{EDL}{electrical double layer}
\nomenclature[zevf]{EVF}{electroviscous flow}
\nomenclature[zfem]{FEM}{finite element method}
\nomenclature[zfvm]{FVM}{finite volume method}
\nomenclature[zpdes]{PDEs}{partial differential equations}
\nomenclature[zsaes]{SAEs}{simultaneous algebraic equations}
%
\nomenclature[aB]{$B$}{charge-dependent slip length (\eqn\ref{eq:11}), --}
\nomenclature[ab]{$b$}{charge-dependent slip length (\eqn\ref{eq:A.16}), m}
\nomenclature[aB0]{$B_\text{0}$}{slip length (\eqn\ref{eq:11}), --}
\nomenclature[ab0]{$b_\text{0}$}{slip length (\eqn\ref{eq:A.16}), m}
\nomenclature[aD]{$\mathcal{D}$}{diffusivity of the positive and negative ions, assumed equal ($\mathcal{D}_{+}=\mathcal{D}_{-}=\mathcal{D}$), m$^2$/s}
\nomenclature[ad]{$d$}{equilibrium distance of Lennard-Jones potential ($=0.4\times 10^{-9}$,  \eqn\ref{eq:A.16}), m}
\nomenclature[adc]{$d_{\text{c}}$}{contraction ratio ($=W_{\text{c}}/W$), --}
\nomenclature[aDj]{$\mathcal{D}_{j}$}{diffusivity of the ions of type j, m$^2$/s}
\nomenclature[ae]{$e$}{elementary charge of a proton ($=1.602176634\times 10^{-19}$), C or A.s}
\nomenclature[aE]{$E_{\text{x}}$}{induced electric field strength (\eqn\ref{eq:A.3}), V/m or --}
\nomenclature[afj]{$\mathbf{f_\text{j}}$}{flux density of the ions of type j (\eqn\ref{eq:A.5}), 1/(m$^2$.s)}
\nomenclature[aIc]{$I_{\text{c}}$}{conduction current density (\eqn\ref{eq:A.11} or \ref{eq:8}), A/m$^2$ or --}
\nomenclature[aId]{$I_{\text{d}}$}{diffusion current density (\eqn\ref{eq:A.11} or \ref{eq:8}), A/m$^2$ or --}
\nomenclature[aIs]{$I_{\text{s}}$}{streaming current density (\eqn\ref{eq:A.11} or \ref{eq:8}), A/m$^2$ or --}
\nomenclature[akB]{$k_{\text{B}}$}{Boltzmann constant ($=1.380649\times 10^{-23}$), J/K}
\nomenclature[alB]{$l_\text{B}$}{Bjerrum length ($=0.7\times 10^{-9}$, \eqn\ref{eq:A.16}), m}
\nomenclature[aLc]{$L_{\text{c}}$}{length of contraction section, m or --}
\nomenclature[aLd]{$L_{\text{d}}$}{length of downstream outlet section, m or --}
\nomenclature[aLu]{$L_{\text{u}}$}{length of upstream inlet section, m or --}
\nomenclature[an+]{$n_{+}$}{local number density of positive ions (\eqn\ref{eq:A.9} or  \ref{eq:6}), 1/m$^3$ or --}
\nomenclature[an-]{$n_{-}$}{local number density of positive ions (\eqn\ref{eq:A.9} or  \ref{eq:6}), 1/m$^3$ or --}
\nomenclature[an0]{$n_{0}$}{bulk (i.e. geometric mean) density of the ions of type j, 1/m$^3$}
\nomenclature[anj]{$n_{j}$}{local number density of the ions of type j, 1/m$^3$}
\nomenclature[ans]{$n^*$}{excess charge ($=n_{+}-n_{-}$), 1/m$^3$ or --}
\nomenclature[aP]{$P$}{pressure, Pa or --}
\nomenclature[aT]{$T$}{temperature, K}
\nomenclature[aU]{$U$}{total electrical potential, V or --}
\nomenclature[aV]{$\mathbf{V}$}{velocity vector, m/s or --}
\nomenclature[aVa]{$\overline{V}$}{average velocity of the fluid at the inlet, m/s}
\nomenclature[aVx]{$V_x$}{x-component of the velocity, m/s or --}
\nomenclature[aVy]{$V_y$}{y-component of the velocity, m/s or --}
\nomenclature[aW]{$W$}{cross-sectional width of inlet and outlet sections, m}
\nomenclature[aWc]{$W_{\text{c}}$}{cross-sectional width of contraction section, m}
\nomenclature[ax]{$x$}{streamwise coordinate, --}
\nomenclature[ay]{$y$}{transverse coordinate, --}
\nomenclature[aY]{$Y$}{electroviscous correction factor (\eqns\ref{eq:15}, \ref{eq:16}, and \ref{eq:24}), --}
\nomenclature[azj]{$z_{j}$}{valency of the ions of type j, assumed equal ($z_{+}=z_{-}=z$), --}
%
%
\nomenclature[gdP]{$\Delta P$}{pressure drop (\eqns\ref{eq:15} and \ref{eq:17}), --}
\nomenclature[geps0]{$\varepsilon_{\text{0}}$}{permittivity of free space (i.e. vaccum), F/m or C/(V.m)}
\nomenclature[gepsr]{$\varepsilon_{\text{r}}$}{dielectric constant (or absolute permittivity or relative permittivity) of the electrolyte liquid, --}
\nomenclature[glambdad]{$\lambda_{\text{D}}$}{Debye length $\left(=\sqrt{\frac{\varepsilon_{\text{0}}\varepsilon_{\text{r}} k_{\text{b}}T}{z^2e^2n_{\text{0}}}}\right)$, m}
\nomenclature[gmu]{$\mu$}{viscosity, Pa.s}
\nomenclature[gmueff]{$\mu_\text{eff}$}{effective or apparent viscosity, Pa.s}
\nomenclature[gpsi]{$\psi$}{EDL potential, V or --}
\nomenclature[grho]{$\rho$}{density of fluid, kg/m$^3$}
\nomenclature[grhoe]{$\rho_{\text{e}}$}{charge density of liquid, C/m$^3$}
\nomenclature[gsigma]{$\sigma$}{surface charge density, C/m$^2$}
\nomenclature[gsigmae]{$\sigma_{\text{e}}$}{electrical conductivity of an electrolyte solution (\eqn\ref{eq:A.12}), A/(V.m)}
%
%
\nomenclature[dbeta]{$\mathit{\beta}$}{liquid parameter (\eqn\ref{eq:5}), --}
\nomenclature[dK]{$\mathit{K}$}{inverse Debye length (\eqn\ref{eq:5}), --}
\nomenclature[dPe]{$Pe$}{Peclet number ($={Re}~\mathit{Sc}$) (\eqn\ref{eq:5}), --}
\nomenclature[dRe]{$Re$}{Reynolds number (\eqn\ref{eq:5}), --}
\nomenclature[dS]{$\mathit{S}$}{surface charge density (\eqn\ref{eq:10}), --}
\nomenclature[dSc]{$\mathit{Sc}$}{Schmidt number (\eqn\ref{eq:5}), --}
%
\nomenclature[sz]{$0$}{without electroviscous effects}
\nomenclature[szz]{$00$}{without electroviscous and slip effects}
\nomenclature[sc]{$c$}{contraction}
\nomenclature[sd]{$d$}{downstream}
\nomenclature[se]{$e$}{extra or excess}
\nomenclature[sm]{$m$}{mathematical}
\nomenclature[ss]{$s$}{statistical}
\nomenclature[su]{$u$}{upstream}
%